\documentclass[acmsmall]{acmart}

\usepackage{amsmath}
\usepackage{array}
\usepackage{graphicx}
\usepackage{makecell}
\usepackage{multirow}
\usepackage{enumitem}
\usepackage{pifont}

\usepackage[utf8]{inputenc}
\usepackage{url}
\usepackage{booktabs}

\usepackage{amssymb}
\usepackage{bbding}
\usepackage{wasysym}
\usepackage{utfsym}
\usepackage{fontawesome}
\usepackage{pdfpages}

\AtBeginDocument{%
  \providecommand\BibTeX{{%
    \normalfont B\kern-0.5em{\scshape i\kern-0.25em b}\kern-0.8em\TeX}}}

\setcopyright{acmcopyright}
\copyrightyear{2023}
\acmYear{2023}
\acmDOI{XXXXXXX.XXXXXXX}

\acmJournal{JACM}
\acmVolume{37}
\acmNumber{4}
\acmArticle{111}
\acmMonth{8}

\begin{document}

\title{Challenges and Remedies to Privacy and Security in AIGC: Exploring the Potential of Privacy Computing, Blockchain, and Beyond}

\author{Chuan Chen}
\email{chenchuan@mail.sysu.edu.cn}
\author{Zhenpeng Wu}
\email{iswuzp@gmail.com}
\author{Yanyi Lai}
\email{laidou2611@163.com}
\author{Wenlin Ou}
\email{wenlin2000@126.com}
\affiliation{%
  \institution{School of Computer Science and Engineering, Sun Yat-sen University}
  \city{Guangzhou}
  \country{China}
  \postcode{510006}
}

\author{Tianchi Liao}
\email{liaotch@mail2.sysu.edu.cn}
\author{Zibin Zheng}
\email{zhzibin@mail.sysu.edu.cn}
\authornote{corresponding author}
\affiliation{%
  \institution{School of Software Engineering, Sun Yat-sen University}
  \city{Zhuhai}
  \country{China}
  \postcode{519000}
}

\renewcommand{\shortauthors}{Chuan Chen and Zhenpeng Wu, et al.}

\begin{abstract}
  Artificial Intelligence Generated Content (AIGC) is one of the latest achievements in AI development. The content generated by related applications, such as text, images and audio, has sparked a heated discussion. Various derived AIGC applications are also gradually entering all walks of life, bringing unimaginable impact to people's daily lives. However, the rapid development of such generative tools has also raised concerns about privacy and security issues, and even copyright issues in AIGC. We note that advanced technologies such as blockchain and privacy computing can be combined with AIGC tools, but no work has yet been done to investigate their relevance and prospect in a systematic and detailed way. Therefore it is necessary to investigate how they can be used to protect the privacy and security of data in AIGC by fully exploring the aforementioned technologies. In this paper, we first systematically review the concept, classification and underlying technologies of AIGC. Then, we discuss the privacy and security challenges faced by AIGC from multiple perspectives and purposefully list the countermeasures that currently exist. We hope our survey will help researchers and industry to build a more secure and robust AIGC system.
\end{abstract}


\keywords{AIGC, Blockchain, Privacy Computing, Copyright, Data Security.}


\maketitle

\section{Introduction}
Artificial intelligence-generated content (AIGC) refers to content generation with advanced generative artificial intelligence (GAI) technology, which can automatically create large amounts of content in a short period of time. This chapter briefly introduces the application of AIGC and its history and emphasizes the importance of discussing data security and privacy protection in the context of AIGC.

\subsection{Background}
AIGC is a novel method for generating images, text, audio and other content using artificial intelligence techniques. AIGC uses techniques such as deep learning, Generative Adversarial Network (GAN) or Variational Auto-encoder (VAE) to learn the latent distribution of data from large-scale datasets and thus generate realistic, high-quality new data.

In recent years, AIGC has developed rapidly, and ChatGPT \cite{chatgpt}, which has attracted widespread attention in 2022, has a strong performance in applications such as human-computer interaction, dialogue, and output generation text. In principle, ChatGPT is a machine learning system based on Large Language Model (LLM) \cite{li2021pretrained}, an LLM with hundreds of billions of parameters. After training on a large number of text datasets, ChatGPT is able to complete more reasonable dialogues and also to create texts with unique styles.

In image generation, such as stable diffusion \cite{stablediffusion}, Midjourney \cite{midjourney} have a good performance, they are the application of Diffusion Model \cite{ho2020denoising,croitoru2023diffusion}, which is given a prompt text, can generate a new image consistent with the requirements of the prompt text. StyleGAN \cite{karras2019style} is also an excellent generative model. It is based on the structure of GAN and is used for image generation, art and design.

The basic steps of AIGC model are the same as pre-training large models \cite{li2021pretrained}. Firstly, collecting data and completing data pre-processing steps. Then the training of the model is completed with the help of the training dataset, the model is fine-tuned according to different requirements, and then the effect of the model is inferred. Finally, the release phase of the model is done to release the pre-trained model. Data security and privacy issues are most likely to occur in the first step of data collection, where model developers may use some unauthorized data to complete the training of the model, leading to data leakage issues.

The development of AIGC cannot be separated from the improvement of computing power and continuous research on deep learning. The history of AIGC development can be roughly divided into three stages.
\begin{itemize}
    \item Early Sprouting Stage: During this phase, scholars have explored generative models as well as computer-generated content. In 1957, the world's first computer-composed musical composition, the Illiac Suite, appeared. In the 1960s, Joseph Weizenbaum wrote the world's first chatbot, Eliza, which was able to respond to people's questions. These were meaningful early attempts at intelligent content generation, but the high cost and difficulty of commercialization in the late 1980s and mid-1990s, and therefore limited capital investment, led to no major achievements in AIGC.
    \item Sedimentation and accumulation stage: In the early 20th century, deep learning developed rapidly, and deep learning algorithms such as Convolutional Neural Networks (CNNs) were proposed to solve many problems such as image classification and image segmentation. At the same time, GPU, CPU and other arithmetic devices are increasingly sophisticated, and the Internet is also in the stage of rapid development, providing a huge amount of data for training various artificial intelligence algorithms.
    \item High-speed development stage: In 2014, GAN \cite{goodfellow2014generative} came out of nowhere to give an efficient and feasible network for generating the content, which again sparked the interest in AIGC. This phase is a continuation of the previous two phases of AIGC development, which was mainly based on pre-training large models. In 2018, NVIDIA released StyleGAN \cite{karras2019style} for automatic image generation, and in 2019 DeepMind released DVD-GAN \cite{clark2019efficient} for continuous video generation. 2021 Open AI launched DALL-E \cite{ramesh2021zeroshot} and updated iterative version DALL-E-2 \cite{ramesh2022hierarchical}, mainly for interactive content generation of text and images. Especially in late 2022, OpenAI released ChatGPT, a chatbot application based on GPT3, which once again sparked widespread discussion and enthusiastic attention.
\end{itemize}

AIGC generates content based on artificial intelligence technology to meet users' requirements, and is a complement to traditional content creation methods such as professionally generated content (PGC) and user-generated content (UGC) 
 \cite{lee2021creators,wu2022ai}. In the PGC era, it takes a long time to produce quality content. In the UGC era, everyone can publish their own creations online as users, but the level of creators varies and the quality of output content is difficult to guarantee. AIGC can overcome the shortage of PGC and UGC in terms of quantity and quality, generating a large amount of content as well as being able to output guaranteed quality content after learning.

\subsection{Urgency of Security and Privacy in AIGC}
Although AIGC models has attracted worldwide attention because of its powerful content generation ability and efficiency, there are still concerns about the actual large-scale use, that is, using AIGC services may cause security or privacy problems. The problems may arise from the AIGC include: leakage of user's input data, model attacks, data crawling and the insecurity of generated content.

\subsubsection{Related Laws and Regulations}
In recent years, countries around the world have put forward laws and regulations related to the use of AIGC and resulting data security and privacy problems. The popularity of Chatgpt marks the vigorous development of AIGC models. The rapid development of AIGC models and its powerful generation ability make countries pay more attention to the data security and privacy problems. The data used to train AIGC models may contain sensitive information, such as personal identity information, medical records and financial information. If this sensitive information is leaked, it will bring great risks to individuals and organizations.

In terms of data security, the EU issued \emph{General Data Protection Regulation} (GDPR) \cite{eu2018generaldatapr} in 2018, which is a new EU regulation on the collection, processing and storage of personal data. In 2019, British Airways was fined nearly 200 million pounds for breaching GDPR by disclosing users' information. GDPR is an important legal basis for protecting the rights of individuals to their own information and ensure that information is not disclosed while being employed. Also, China has successively promulgated \emph{the Data Security Law}, \emph{the Personal Information Protection Law}, \emph{the Cybersecurity Law} to ensure the security of personal data in the process of network circulation. In 2022, China issued 20 Articles of Regulation on Data, proposing to build a data-based system to better play the role of data elements, and standardize the correct use of data as elements.

As for AIGC, countries have noticed its powerful generating ability, so they put forward restrictions on its use. In 2018, China Academy of Information and Communications Technology issued WHITE PAPER On AIGC, affirming that AIGC is an indispensable supporting force for moving towards a new era of digital civilization, and giving suggestions on development of AIGC from government, industry, enterprises, and society. In 2023, China Internet Information Office issued \emph{the Notice on Public Solicitation of Opinions on the Management Measures for Generative Artificial Intelligence Services (Draft for Soliciting Opinions)}, which proposed to regulate the use of AIGC algorithm in all aspects, protect users' rights and data security, and develop AI technology with caution.

\subsubsection{Technical Point of View}
In addition to the relevant laws and regulations published by each country, from the technical point of view, there are many privacy and security issues that may arise from the AIGC, and the possibility of privacy security being jeopardized continues to increase. Thus, there is an urgent need to propose solutions to protect data privacy security under AIGC.

\begin{itemize}
    \item Deep fake technology: The rapid growth of AIGC has raised concerns about deep fake technology \cite{wester2019deepfake}, which uses AI-based techniques to generate near-real photographs, movies or audio, which can be utilized to portray events or individuals that do not exist. The emergence of deep fake technology makes it possible to tamper with or generate highly realistic and indistinguishable audio and video content, which ultimately cannot be distinguished by the naked eye of the observer. On the one hand, falsified content can mislead others to believe in the generated fake information, which, if not prevented, will certainly bring new risks to national security, citizens' and enterprises' rights and interests, and cause potential crises in social development. On the other hand, forged content poses a potential threat to personal privacy and security, for example, malicious users can use these images to commit fraud and other illegal activities.
    \item Not good enough content: Not good enough content means content that is not real enough or even toxic. When AI-generated content is not realistic enough, it is easy to make users think that the AIGC-generating ability is limited and have a negative impression on the AIGC model itself, which hinders the development of the AIGC model. When AI-generated content is toxic, it may have an impact on human cognition, which involves ethical and moral issues. Ethics is an aspect that cannot be ignored in AIGC technology development and involves issues such as values, morality, and legal concepts between AI and human society. The potential toxicity of AI-generated content refers to the problem that AI-generated content is biased, i.e., AIGC may generate content that violates social values, so it can easily become a tool for many ill-intentioned people.
    \item Vulnerable models: A malicious user can exploit vulnerabilities in AIGC models to attack models and add intentional interference signals to the input data to deceive the behavior of AIGC models. This may result in the model generating incorrect outputs or generating misleading information in an intentional manner. And, the use of model backpropagation attacks may also infer the data used for the original training of the model from some of the output, which can cause privacy breaches and raise data security issues.
\end{itemize}

\subsection{Contributions}

In this survey, we provide an introduction of AIGC, the security and privacy problem in AIGC and our proposed solution. The contributions of this survey are as follows:

\begin{itemize}
    \item We review and summarize the definition and classification of AIGC, and enumerate its background technologies, basic models and applications, etc. We emphasize the urgency of addressing privacy and security issues in this area.
    \item We analyze the privacy and security challenges faced by AIGC from the perspectives of privacy of circulating data, security of generated content, and threats from malicious users. We illustrate the significant threats to user security posed by these challenges in AIGC by employing specific examples, with a particular focus on copyright issues in AIGC.
    \item We comprehensively conclude the current status of privacy protection and data security in AIGC, and investigate the various response strategies available, with the advantages and drawbacks of these approaches.
    \item We provide the first systematic discussion of the vital role of blockchain and privacy computing in improving the security of AIGC, delivering illustrative cases where technologies such as blockchain, federated learning and digital watermarking are combined with AIGC, and revealing their great potential in addressing AIGC data privacy, content security and intellectual property issues.
\end{itemize}

\begin{table*}[h]
\centering
\renewcommand{\arraystretch}{1.3}
\caption{Summary of related works versus our survey}
\resizebox{\columnwidth}{!}{
\begin{tabular}{|l|l|l|l|l|l|l|l|l|}
\hline
\textbf{Year}  & \textbf{Ref.} & \textbf{Brief contribution} & \textbf{AIGC}  & \textbf{\begin{tabular}[c]{@{}l@{}}Confiden- \\ tiality\end{tabular}}   & \textbf{Copyright}  &\textbf{Blockchain} &\textbf{Watermark}  &\textbf{\begin{tabular}[c]{@{}l@{}}Privacy \\ Computing\end{tabular}} \\ \hline
    \multirow{2}{*}{2019}     & \cite{Sangeetha2019}  & \begin{tabular}[c] 
                               {@{}l@{}}Introduce state-of-art privacy- 
                               protecting data \\ mining mechanisms and their application in big \\ data environment\end{tabular} & \XSolidBrush  &\CheckmarkBold & \XSolidBrush &\XSolidBrush &\XSolidBrush &\CheckmarkBold\\ \cline{2-9} 
                              & \cite{sun2019privacy} & \begin{tabular}[c]{@{}l@{}}Review and analyze cloud computing privacy \\ protection research solutions, and gives research \\challenges and future directions\end{tabular} & \XSolidBrush &\CheckmarkBold  & \XSolidBrush & \XSolidBrush &\XSolidBrush &\CheckmarkBold
                                \\ \hline
\multirow{3}{*}{2020}        & \cite{wang2020privacysurvey}   & \begin{tabular}[c] 
                             {@{}l@{}} Systematically analyze and 
                             summarize the privacy \\ protection based on blockchain\end{tabular}    & \XSolidBrush   
                             &\CheckmarkBold   &\XSolidBrush &\CheckmarkBold &\XSolidBrush &\CheckmarkBold \\ \cline{2-9} 
                             & \cite{qureshi2020block}  & \begin{tabular}[c]{@{}l@{}} Comprehensively summarize the use of blockchain \\ for privacy,  security and copyright protection of \\ multimedia content \end{tabular}  &\XSolidBrush &  \CheckmarkBold &\XSolidBrush &\XSolidBrush &\XSolidBrush &\CheckmarkBold \\ \cline{2-9}
                             & \cite{MOTHUKURI2021619}  &  \begin{tabular}[c]{@{}l@{}} Provide a comprehensive study of the security \\ and privacy  aspects of federated learning \end{tabular}  
                             & \XSolidBrush &\CheckmarkBold &\CheckmarkBold &\CheckmarkBold  &\XSolidBrush &\CheckmarkBold
                             \\ \hline
\multirow{3}{*}{2021}        & \cite{yin2021privacyfl}  & \begin{tabular}[c]{@{}l@{}} 
                            Summarize the privacy protection 
                             methods \\ in the field  of Federated Learning \end{tabular}    & \XSolidBrush &\CheckmarkBold &\XSolidBrush &\XSolidBrush &\XSolidBrush &\CheckmarkBold
                             \\ \cline{2-9} 
                             & \cite{liu2020machine}  & \begin{tabular}[c]{@{}l@{}} Summarize privacy attacks and protection \\ techniques in  machine learning \end{tabular}   &\XSolidBrush &\CheckmarkBold &\XSolidBrush &\XSolidBrush
                             &\XSolidBrush &\CheckmarkBold
                             \\ \cline{2-9}
                             & \cite{byrnes2021data} & \begin{tabular} 
                             [c]{@{}l@{}} Discuss how digital watermarking techniques can \\ be applied to deep learning networks to facilitate \\ the development of responsible AI \end{tabular} &\XSolidBrush  & \XSolidBrush & \XSolidBrush &\XSolidBrush &\CheckmarkBold &\XSolidBrush
                              \\ \hline
\multirow{2}{*}{2022}        & \cite{WAN2022226} & \begin{tabular}[c]{@{}l@{}} 
                             Summarize the various existing image \\ watermarking methods
                             \end{tabular} &\XSolidBrush  & \XSolidBrush & \CheckmarkBold &\XSolidBrush &\CheckmarkBold &\XSolidBrush
                             \\ \cline{2-9} 
                             & \cite{wadhera2022comprehensive} &\begin{tabular} 
                             [c]{@{}l@{}} Discuss the specific application scenarios \\ of digital watermarking\end{tabular}&\XSolidBrush  & \XSolidBrush & \CheckmarkBold &\XSolidBrush &\CheckmarkBold &\XSolidBrush
                             \\ \hline
\multirow{5}{*}{2023}         & \cite{cao2023comprehensive} & \begin{tabular}[c] 
                              {@{}l@{}} Provide a official definition and 
                              comprehensive \\ survey of  the generation process for AIGC and \\ AI enhancements\end{tabular} &\CheckmarkBold &\CheckmarkBold &\XSolidBrush &\XSolidBrush &\XSolidBrush &\CheckmarkBold
                             \\ \cline{2-9}
                             & \cite{wu2023aigenerated} & \begin{tabular}[c]{@{}l@{}} Provide a broad overview of AIGC, including \\ its definition, basic conditions, cutting-edge \\ capabilities,\\ advanced features and AIGC industry chain \end{tabular} &\CheckmarkBold &\XSolidBrush &\XSolidBrush &\XSolidBrush &\XSolidBrush &\XSolidBrush
                             \\ \cline{2-9}
                             & \cite{xu2023unleashing} & \begin{tabular}[c]{@{}l@{}} Describe the applications and challenges of \\ mobile AIGC network deployment\end{tabular} &\CheckmarkBold &\CheckmarkBold &\XSolidBrush &\CheckmarkBold
                             &\XSolidBrush &\CheckmarkBold
                             \\ \cline{2-9}
                             & \cite{chen2023pathway} & \begin{tabular}[c]{@{}l@{}}Discuss the privacy security and copyright issues \\ of AI-generated content\end{tabular} &\CheckmarkBold &\CheckmarkBold &\CheckmarkBold &\XSolidBrush &\XSolidBrush &\CheckmarkBold
                             \\ \cline{2-9}
                             & \begin{tabular}[c]{@{}l@{}}Ours\end{tabular} & \begin{tabular}[c]{@{}l@{}} Summarize the existing AIGC technology, analyze \\ the privacy and security challenges faced by AIGC, \\ summarize the current status of AIGC in data \\ security  and privacy protection, and  finally give \\ concrete examples of the combination of  blockchain, \\ federated learning, and digital watermarking \\ technologies with AIGC \end{tabular}&\CheckmarkBold &\CheckmarkBold &\CheckmarkBold &\CheckmarkBold &\CheckmarkBold &\CheckmarkBold
                             \\ \hline
\end{tabular}}
\end{table*}

\section{Preliminary and Relevant Technology}
In this section, the background and details about AIGC technology are presented. Specifically, we examine the definition of AIGC, its popular model and application, and the privacy and security technology that could be used in AIGC.

\subsection{Definition of AIGC}
In the development of the Internet, there are three stages of content creation: PGC, UGC and AIGC. Below we will give the detailed definitions and distinctions of them.

PGC, Professionally generated Content, is a mode of producing highly professional content. In the Early Internet, most of the websites such as video broadcasting websites adopted PGC mode \cite{tiago2019youtube}. Most of the content was created by professional practitioners, so the quality of content could be guaranteed and users had a good experience of reading content. Consequently, PGC relied on its own professionalism and quality content to retain users. However, professional production requires a large number of experts, and users often adopt extremely high standards for content, that is, PGC needs to ensure that content is as high quality as ever in order not to lose users.

UGC, User-generated Content, is a mode that digital content is generated by users rather than professionals \cite{krumm2008user}. UGC makes it possible for users to create content at their own pace, posting original videos and graphics on social media platforms such as Twitter, YouTuBe, etc. The UGC lowers the threshold for users to create content, and the wide user base makes content more creative and diverse, UGC is more naive than PGC.

AIGC, AI-generated Content, is the content output by the user's input through generative models. In recent years, AIGC has begun to be considered as a new type of content creation alongside PGC and UGC. AIGC can learn the underlying artistic style and creative patterns from the countless work created by humans, so AIGC models can generate more diverse and realistic content based on it. Recently, the mighty ability of stable diffusion based on the diffusion model in image generation and ChatGPT based on the transformer model in text generation has made AIGC gain great attention.

Compared with PGC and UGC, AIGC has higher efficiency and diversity. The most immediate advantages brought by AIGC are speed, creativity and more possibilities, for example, AI painting can help creators broaden their ideas and speed up their creative efficiency. These advantages allow AIGC to be applied to various fields. Further, we define the characteristics of AIGC as follows

\begin{itemize}
    \item Efficient: Due to the continuous development of deep learning and many accelerated training methods proposed in recent years, AIGC has become very efficient in generating content \cite{cao2023comprehensive}. When you chat or ask a question to ChatGPT, the generated result is presented to the user by ChatGPT in less than a minute. Stable diffusion, as a representative of image generation technology, takes about two to three minutes to generate a new image that is semantically consistent with the input prompt. The efficiency of AIGC provides unparalleled benefits for using artificial intelligence technologies to generate content.
    \item Creative: Because AIGC learns from a wide variety of training datasets, so it is able to acquire different styles and characteristics from creators and combine them organically in the output, ultimately creating a refreshing creation \cite{zhang2023completeaigc}. And it has been noticed in the extensive use of AIGC in reality that it is indeed very creative and often brings a novel experience to people.
    \item Diverse: The training data of AIGC models comes from a large amount of data from many aspects. Due to this large amount of data from different fields, AIGC is not limited to narrow topics, for example, stable diffusion can generate cat pictures not just a same cat or a same breed of cat. The diversity of AIGC brings more possibilities for creation and the content will not become boring.
    \item General: AIGC applications come from pre-trained large models, and the power of large models makes it effective for a wide variety of downstream tasks \cite{gao2023chatrec}. For example, ChatGPT not only excels in text creation in various styles, but also shows notable results in question answering, and even assists in code modification, all based on the excellent generality of its core GPT.
\end{itemize}

\subsection{General Techniques of AIGC}
In this section, we will introduce the basic techniques behind AIGC models. The introduction of general techniques helps to gain a deeper understanding of the operating principles of AIGC.
\subsubsection{Recurrent Neural Networks}
Recurrent Neural Networks (RNN) \cite{grave2013rnn} is an improvement of feed-forward neural networks. Feed-forward neural networks do not take into account the correlation between data, and the output of network is only related to the input of network at the current moment. However, when solving real-world problems, it is found that there are many sequential data, such as text, voice and video, which are often time-series correlated. Thus, RNN was proposed to exploit these correlations well. Feed-forward neural network has three-layer structure: input layer, hidden layer and output layer. Generally, the output at moment t comes from the input at moment $t$ though hidden layer. RNN adds an information flow that the hidden at moment t is related to hidden at moment $t-1$. 

When the sequence becomes longer, however, there is a problem of exploding or disappearing gradients. To reduce long-term dependence and eliminate the effects of gradient disappearance, Sepp  \emph{et al.} \cite{hochr19997lstm} proposed Long Short-Term Memory (LSTM) model, it is designed to store and update information by designing three gates: an input gate, an oblivion gate and an output gate. By merging the unitary and hidden states of the LSTM, Gate Recurrent Unit (GRU) \cite{chung2014empirical} was proposed to simplifies the LSTM by replacing the forgetting gate and the input gate with the update state. RNNs can have various architectures with different numbers of inputs and outputs:one-to-one, many-to-one, one-to-many, and many-to-many. Many-to-many can be used for machine translation, also known as sequence-to-sequence (seq2seq) model \cite{NIPS2014_a14ac55a,cho2014learning}.
\subsubsection{Convolutional Neural Networks}
Convolutional neural networks (CNN) \cite{wai1989phoneme,le2010convo} have long been the standard backbone in computer vision, the core of which lies in the design of its convolutional layers, which not only share parameter weights but also are able to extract features.

VGG \cite{simonyan2015deep} demonstrated that deeper network layers are an effective means to improve accuracy, but deeper networks are highly susceptible to gradient dispersion, which leads to network failure to converge. It has been tested that more than 20 layers will converge less and less effectively as the number of layers increases. Previously, when deeper layers are used in deep learning, the training results tend to become extremely poor, so researchers need to repeatedly and continuously debug to select the optimal number of layers for the network.

Resnet \cite{He2016CVPR} is an important development of CNN that introduces residual connectivity and stabilizes the training of deep learning with deeper layers. Since the introduction of Resnet, deep learning can take on an infinitely deep network structure without performance deteriorating as the network becomes too deep. DenseNet \cite{Huang2017CVPR} significantly reduces the number of parameters in the network by feature reuse, which in turn alleviates the gradient disappearance problem to some extent.
\subsubsection{Transformer}
In 2017, Transformer \cite{NIPS2017_3f5ee243} was proposed, which uses a Self-Attention structure and gradually replaces the RNN network structure commonly used in natural language processing tasks.The biggest advantage over the RNN network structure is that it can be computed in parallel.

Transformer uses an Encoder-Decoder architecture, and the tasks it needs to perform are shown in the figure:

Each encoder consists of two sub-blocks: the Self-Attention layer and the Position-wise Feed Forward Network. the core of the transformer is the self-attention layer. Transformer has become a well-established standard approach in natural language processing tasks.
\subsubsection{ViT}
Inspired by the success of the Transformer in the natural language processing field, many works have attempted to apply the Transformer to computer vision field \cite{dosovitskiy2021image,touvron2021training,carion2020endtoend}, which is called Vision Transformer (ViT). \cite{dosovitskiy2021image} first flatten the image into a series of two-dimensional chunks and inserts a class token at the beginning of the sequence to extract the classification information. After encoding the embedding position, the marker embedding is fed into the standard Transformer. This simple and effective implementation of ViT makes it highly scalable. It argues that ViT outperforms CNN when having enough data for pre-training, breaking the limitation of the Transformer's lack of inductive bias and allowing for better migration in downstream tasks. DeiT\cite{touvron2021training} a teacher-student strategy for training to reduce Transformer's reliance on large data by introducing distillation tokens. DETR \cite{carion2020endtoend} applies Transformer to object detection, replacing the current work of models that require manual design, and achieves good results. In the current paradigm of large models and large data sets, Transformer has gradually replaced CNN as the dominant model in computer vision.

The four backbones introduced above are the basic structure of the AIGC model.

\subsection{Classical Model of AIGC}
In this section, we mainly introduce five classical AI-generated models. They give various ways to generate content, and we can choose different models according to different requirements.
\subsubsection{Based on generative adversarial networks}
The proposal of GAN \cite{goodfellow2014generative} marks the continued development of AIGC in a new era, including both generative and discriminative models. Generative adversarial networks employ the idea of mutual games between generative and discriminative models. The generative model is based on the original real data to generate a better quality image that is closer to the real data distribution and allows this image to deceive the discriminant model. The task of the discriminative model is to distinguish the real data from the false data generated by the generative model. In the training process of GAN, the generative model and the discriminant model influence each other and compete with each other so that the generator can generate better quality images and achieve a stable state.

The generative power of GAN is powerful but still has many drawbacks, such as mode collapse, where the generator easily falls into model collapse and locks a single pattern to trick the discriminator.

\subsubsection{Based on Diffusion Model}
Unlike the GAN that games between the generative and discriminative models, the Diffusion model \cite{ho2020denoising} trains the inverse process with the assumption of the forward process to obtain the final image output. Specifically, the diffusion model contains a forward process and an inverse process. The forward process assumes that the image at moment $t$ is obtained from the image at moment $t-1$ plus Gaussian noise, which can be considered as a Markov process. The inverse process recovers the original image $x_0$ in reverse from the image at moment $t$ with $t$ rounds of Gaussian noise added (when $t$ is large, the image can be considered as a complete Gaussian noise), and the diffusion model obtains this inverse process with the help of deep neural network training, thus gaining the ability to recover the original image $x_0$ from complete noise.

Stable Diffusion \cite{stablediffusion} is a STAR variant of diffusion model, which accomplishes such forward and backward processes in latent space, reducing the time and content needed for model training and speeding up the efficiency. The original stable diffusion generates better results on images with 512 $\times$ 512 resolution, and Dist-diffusion \cite{meng2023distillation} also generates good results on low resolution 256 $ \times$ 256 images after introducing the teacher-student strategy.

\subsubsection{Energy-based Generative model}
Energy-based models optimize the model by constructing an energy function and by minimizing this energy function with respect to the input features. EBGAN \cite{zhao2017energybased} changes the discriminator to an energy function that will assign low energy values to regions with high data density and higher energy values to other regions. Due to the introduction of the energy function, the control over the quality and diversity of the generated samples is more stable and can produce a richer and more realistic set of samples.
\subsubsection{Flow-based Generative model}
The stream-based generative model is actually a generator in which a probability density function is defined. The approach of the stream-based generative model is to make the probability density function of the generator infinitely close to the probability density function of the real sample distribution, which simplifies the generative model using the formula of the probability flow \cite{pmlr-v97-ho19a} and improves the training and learning efficiency.
\subsubsection{Variational Auto-Encoder}
Variational Auto-Encoder (VAE) is one of the deep generative models, proposed by Kingma  \emph{et al.} \cite{MAL-056} in 2014. Unlike the traditional self-encoder that describes the latent space by numerical means, it makes observations of the latent space in a probabilistic way, which is of high value for applications in data generation.
VAE is divided into two parts, the encoder and the decoder. The encoder converts the original high-dimensional input data into a probabilistic distribution description of the latent space, and the decoder generates new data by reconstruction from the sampled data.

\subsection{Applications of AIGC}
AIGC has many applications, and the following will briefly introduce a few of its more widely used areas, including the research algorithms and applications in the field.
\subsubsection{Text Generation}
ChatGPT \cite{chatgpt} is the star application since recently, and the latest advancement of the technology makes the artificially generated text almost indistinguishable from the human-written text with stunning results. Bert \cite{devlin2019bert} is also a language model based on Transformer. It is only necessary to insert the input and output of a specific task into Bert, and many downstream tasks can be simulated by using the powerful attention mechanism of Transformer.
\subsubsection{Image Generation}
The proposal of GAN makes it possible to generate images of more consistent quality.  StyleGAN \cite{Karras_2019_CVPR} is a new generative network proposed by NVIDIA after ProGAN \cite{karras2018progressive}, which focuses on controlling the visual features represented in each layer by modifying the inputs of that layer separately, without affecting the other layers. These features can be coarse features (e.g., pose, face shape, etc.) or some detailed features (e.g., pupil color, hair color, etc.) The EditGAN \cite{ling2021editgan} allows users to modify the detailed object parts, thus enabling high precision semantic image editing.
\subsubsection{Video Generation}
In image generation, existing AIGC techniques have been able to generate images with such realism that even the naked eye cannot distinguish the authenticity of the generated image, and it is hoped that this effect can be achieved in video generation as well. DeepMind has proposed DVD-GAN \cite{clark2019efficient}, which can be extended to longer and higher resolution videos using computationally efficient discriminator decomposition. DVD-GAN is capable of generating a continuous video, and this research is an exploration towards realistic video generation. Video diffusion \cite{hovideo} was the first to use the Diffusion model for large-scale video generation and has produced very good video results on several well-known datasets, both conditionally and unconditionally.
\subsubsection{Voice Generation}
AI-generated audio is a valuable area of research because of its potential to enhance the user experience and improve efficiency. WaveNet \cite{vanwavenet}, an artificial intelligence developed by DeepMind, is the first deep neural network capable of generating natural human speech. This technology is able to generate relatively realistic-sounding human-like voices by directly simulating waveforms using a neural network approach trained with real voice recordings.
\subsubsection{3D Scenes Generation}
3D scene generation can be seen as a novel view synthesis task. The more popular model recently is NeRF \cite{mildenhall2021nerf}, it uses a 5-dimensional neural radiance field as an implicit representation of complex 3D scenes and proposes a micro-renderable process based on the classical Volume rendering. Block-nerf \cite{tancik2022block} can render large scenes, such as a 3D scene of a city street scene. ARF \cite{zhang2022arf} can do style migration on 3D scenes, i.e. render images with different views with characteristic styles and maintain consistency between views. Magic3D \cite{lin2023magic3d} can do the job of generating from text to 3D scenes.

\subsection{Privacy and Security Technology}

In the era of Big Data, data security and privacy protection issues are inescapable, especially under the prevalence of AIGC, such issues will be more serious. This section will briefly introduce four related technologies that can be used to solve the data security and privacy protection problems in AIGC.

\begin{figure*}
	\centering
	\includegraphics[width=1\textwidth]{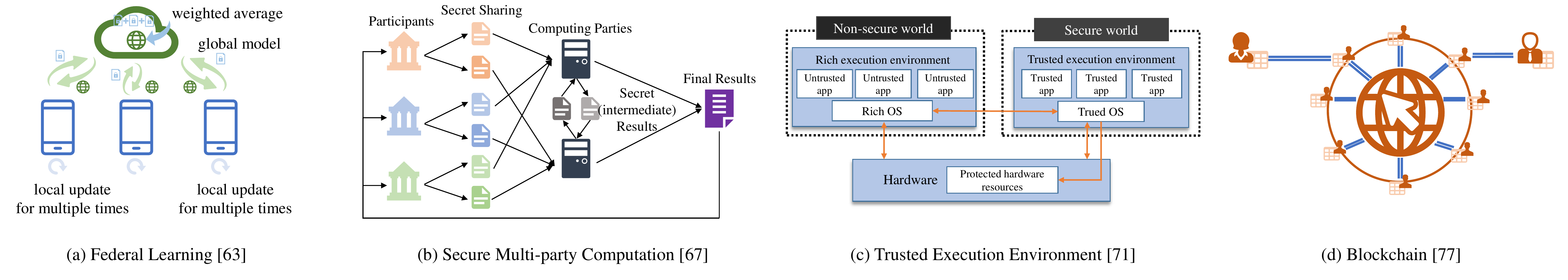}
	\caption{Privacy and security technology.}
	\label{fig: technology}
\end{figure*}

\subsubsection{Federated Learning}
Federated Learning \cite{mcmahan2017communication,li2020review,hamer2020fedboost} is able to consider privacy protection and data security issues when training AI-generative models. Compared with traditional machine learning using a centralized approach to training, federated learning achieves joint multi-party learning training by circulating and processing intermediate encrypted data without local raw data out of the library.

FedAvg \cite{mcmahan2017communication} proposes a distributed framework that allows many users to train a model simultaneously. There is no need to upload any private data to the server during the training process, effectively reducing the privacy risk associated with data aggregation from traditional machine learning sources. Limited by communication costs, large models are difficult to be directly applied in federation learning. Fedboost \cite{hamer2020fedboost} addresses the above challenges by integrating learning and demonstrates the convergence of the proposed federation integration method.

Federated learning has been applied in many ways \cite{li2020review,li2021unified,li2022decentralized}. For example, it can be extended to include enterprises across organizations in a federation framework. A bank with data on customer purchasing power can collaborate with an e-commerce platform with data on product characteristics to recommend products.
\subsubsection{Secure Multi-party Computation}
Secure Multi-party Computation(SMC) \cite{goldreich1998secure,du2001secure,torkzadehmahani2022privacy} is a privacy-preserving distributed computing technique in cryptography. In other words, SMC technology can capture the value of data usage without revealing the original data content.

In recent years, many studies have explored the application of SMC techniques in deep learning. To facilitate the adoption of SMC in machine learning, \cite{knott2021crypten} proposed CRYPTEN, a software framework that builds SMC frameworks in deep learning by means of computations commonly found in modern machine learning frameworks, such as tensor calculus, automatic differentiation, and modular neural networks. \cite{tran2021efficient} developed a framework called secure decentralized training framework that does not require trusted third-party servers and can ensure the privacy of local data with low communication cost. An efficient secure and protocol is proposed that can jointly compute the sum of private inputs and can handle both integers and floating point numbers without any data transformation.
\subsubsection{Trusted Execution Environment}
Trusted Execution Environment (TEE) \cite{jauernig2020trusted,teepicture} refers to an isolated and secure execution environment where programs and data are protected at a higher level of security than at the operating system level. TEE is based on hardware-secure CPU that enables memory-isolated secure computing, allowing privacy-protected computing while maintaining computational efficiency. Sabt  \emph{et al.} \cite{sabt2015trusted} uses the separation kernel to propose a precise definition of TEE and analyze its core properties. Keystone \cite{lee2020keystone} uses simple abstractions provided by hardware, such as memory isolation and programmable layers under untrusted components (e.g., operating systems).
\subsubsection{Blockchain}
Blockchain \cite{nakamoto2008bitcoin,zheng2018blockchain,tschorsch2016bitcoin,blockchainpicture} is a decentralized distributed database technology, which guarantees the reliability and security of transactions through consensus algorithms and cryptography among network nodes. Each transaction must be verified and validated by multiple nodes before it can be written to Blockchain. 

As a decentralized and secure transaction setup, Blockchain has been used in many fields, such as finance \cite{treleaven2017blockchain}, intellectual property protection \cite{song2021proof}, and supply chain management \cite{queiroz2020blockchain}. Therefore, due to its distributed and security features, it provides higher data security and trustworthiness, allowing Blockchain to be applied in the AIGC model to protect data security and privacy protection.

\section{Challenges}
\begin{figure*}
	\centering
	\includegraphics[width=0.8\textwidth]{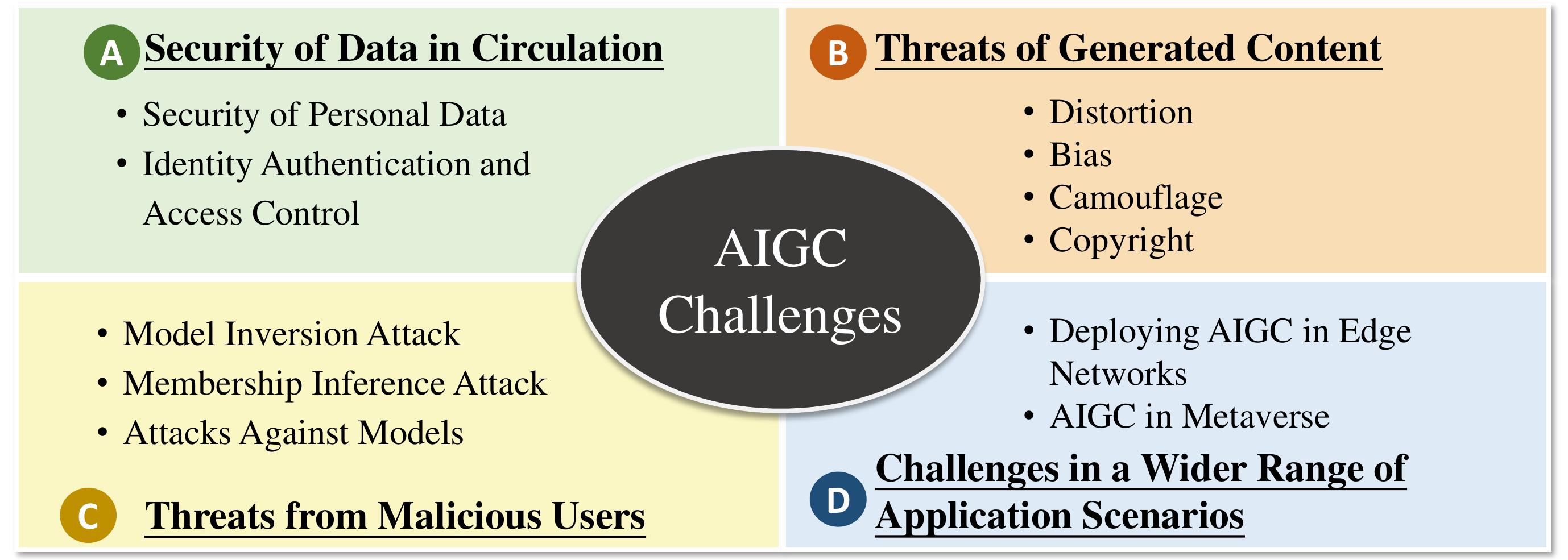}
	\caption{Challenges of AIGC.}
	\label{fig: challenges}
\end{figure*}

From the previous section, we can know that AIGC, as an emerging content production method, is being increasingly used in various fields. Nevertheless, as shown in Fig. \ref{fig: challenges}, there are some privacy and security challenges in AIGC. Generative AI models may employ user data as training data for further iterations, which raises significant concerns about the privacy of user data. In addition, the content generated by generative AI models is difficult to control, which may contain falsified and deceptive content that can give incorrect guidance to users, or discriminatory and biased content that may cause adverse social impact. Generative AI models themselves may also be attacked by malicious users, leading to some privacy and security issues. In this section, we will introduce the privacy and security challenges in AIGC, which are important issues that cannot be ignored in the practical application of AIGC.

\subsection{Security of Data in Circulation}

When using AIGC services, users will inevitably upload some of their personal data to AIGC servers, and there are some security risks in the process of data circulation.

\subsubsection{Security of Personal Data}

Uploading personal sensitive data directly to generative AI models is a risky practice. AIGC services typically use large-scale generative AI models, while Plant \emph{et al.} \cite{plant2022you} showed a positive correlation between the complexity of large language models, the amount of data used in pre-training, and data leakage, implying that AIGC has a higher risk of data leakage. The reason for this problem is that AIGC models usually have a large number of parameters and a high ability to learn quickly with a small number of training samples \cite{brown2020language}. This means that the model is able to learn the personal information provided by the user, such as name, phone number, and address, after a short interaction with the user. Currently, there is no sufficiently effective means to protect users' personal data from being violated. In recently popular ChatGPT, it is technically almost impossible for the development company OpenAI to remove all users' personal information from the data provided to ChatGPT at present, despite the fact that OpenAI claims to remove it from the data before using it.

\begin{figure*}
	\centering
	\includegraphics[width=0.8\textwidth]{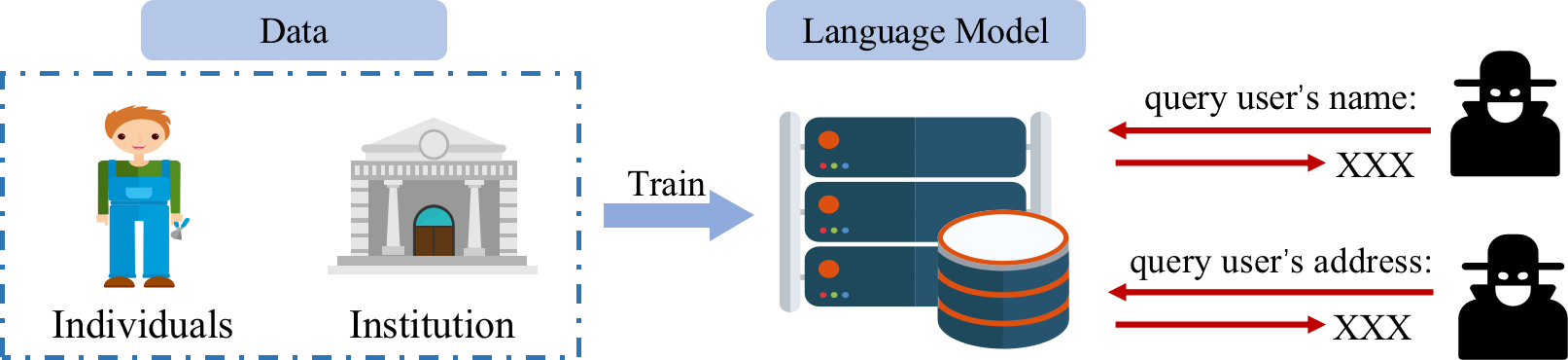}
	\caption{Data leakage in the large language model.}
	\label{fig: leakage}
\end{figure*}

Another important reason for user privacy leaks is that the model itself is not equipped to protect private information. As illustrated in Fig. \ref{fig: leakage}, malicious users may obtain privacy information from other users by querying large language models. If the data provided by users to the AIGC model is used as training data for further iterations of the model, the private data will be recorded in the model as model weights. In the process of interacting with users, the AIGC model may disclose the private data of other users \cite{emmery2021adversarial}.

The inability to guarantee the security of users’ personal data is an important problem faced by AIGC. The solution to this problem can be considered from several aspects. This includes identifying and removing private information from user-provided data, protecting the original user data by some privacy security methods and preventing the model from outputting content related to private data. However, currently, AIGC is still unable to handle users' personal data well, which is a challenge for AIGC in the foreseeable future.

\subsubsection{Identity Authentication and Access Control}

The powerful capabilities of large generative AI models enable them to quickly learn user privacy data. However, presenting this data to all users without reservation by AIGC services poses serious security issues. The purpose of identity authentication and access control is to restrict users with different identities from accessing specific data. However, there is currently a lack of corresponding restriction measures in AIGC services. Companies including Microsoft and Amazon have warned their employees not to share internal confidential information with ChatGPT, as there have been instances where the output of ChatGPT closely matches existing confidential material.

Currently, the development of authentication and permission control technology in AIGC is still in its infancy. Although authentication and permission control technologies are relatively mature in traditional Web applications, supporting authentication by passwords \cite{papathanasaki2022modern}, tokens \cite{el2019survey}, biometrics \cite{rui2018survey}, etc., the situation is much more complex in the AIGC scenario, which makes the current technologies not yet fully adaptable to the AIGC scenario. Firstly, since AIGC is trained based on massive amounts of data, and the data provided by users often has a high degree of randomness and diversity, it is a great challenge to accurately classify and process such massive data to support data access control. Secondly, the large number of users requires highly efficient identity authentication algorithms to ensure that users can use AIGC services normally. In addition, some existing identity authentication technologies, such as biometric based identity authentication, require users to provide more privacy information, which further increases the risk of privacy leakage and may have counterproductive effects.

The special properties of AIGC, such as massive data and high dynamism, make traditional identity authentication and access control technologies not fully applicable. To improve the privacy and security performance of AIGC, improvements need to be made to address these characteristics, considering how to achieve efficient identity authentication and fine-grained access control in massive amounts of users and data.

\subsection{Threats of Generated Content}

AI-generated content has a certain degree of randomness, and humans cannot fully control AI-generated content. Therefore, AI-generated content has security risks and may have some impacts on human society. Specifically, the threats of AI-generated content can be summarized into four aspects: distortion, bias, camouflage, and copyright.

\subsubsection{Distortion}

The distortion of AI-generated content refers to content that is contrary to the facts, generating false information and misleading users. Such content affects the accuracy of the information and may negatively affect the user's decision making.

Bender \emph{et al.} \cite{bender2021dangers} points out that large language models generate content that may be meaningless and untrue, and in some specific domains, this factually contrary information may have a serious impact \cite{allen2017artificial}. In the field of journalism, fake news may be generated if the AIGC model misinterprets news facts. Brundage \emph{et al.} \cite{brundage2018malicious} points out that the application of AIGC increases the likelihood of fake news generation and that AIGC also promotes the mass dissemination of fake news \cite{aimeur2023fake}. The widespread fake news may cause unnecessary public panic and pose a threat to the harmony of society. In addition, fake news may smear the image of enterprises or individuals, which directly causes economic losses to enterprises and reputation problems to individuals. In the medical field, the distortion of AI-generated content is extremely harmful. Medical diagnosis needs to be based on accurate information, and if AI-generated diagnostic suggestions are against the facts, it will cause irreparable harm to patients' lives and health. In the field of e-commerce, fake advertisements can mislead consumers' purchasing decisions and lead to damage to consumers' interests, while fake reviews generated by AI may both mislead consumers and adversely affect the reputation of enterprises.

Another threat lurking in AI generated content is that AI models often interpret the generated content. When the AI model cannot answer the user's question correctly, it will still output some patchwork information regardless of the correctness of the information. AIGC will not only provide wrong information but also make seemingly reasonable defenses for it, accompanied by indistinguishable arguments that are more confusing, which makes the AI-generated content that contradicts the facts easier for users to trust.

\subsubsection{Bias}

The biases of AI generated content include inconsistency with human values, stereotypes or discrimination towards specific groups, which can harm social harmony and exacerbate conflicts between different groups.

It has been demonstrated that the AIGC model as an agent for humans, such as the large language model as a language agent for humans, is not entirely consistent with human intentions, meaning that the AI model may not act entirely according to the designer's intentions \cite{gabriel2020artificial,russell2019human}. This issue is called the "AI alignment problem" and has undergone some research in the past. However, the enhancement of AI model capabilities makes it increasingly difficult for humans to intervene in this \cite{kenton2021alignment}, and the powerful capabilities of AIGC models make this issue even more apparent.

Due to the fact that the training of the AIGC model is based on massive data, without intervention measures, bias in the training data will lead to corresponding bias in the content generated by AI. These biases involve various themes, such as religious stereotypes, gender bias, racism, etc. Abid \emph{et al.} \cite{abid2021persistent} demonstrated through experiments that the advanced language model GPT-3 has a sustained violent bias against Muslim communities. The author tests GPT-3 through various tasks, including analogical reasoning and story generation, and found that the model has a significantly higher violent bias against Muslim communities compared to other religious groups. Sheng \emph{et al.} \cite{sheng2019woman} found that language models have gender bias, suggesting that the association between women and profession is smaller than that of men, indicating that AIGC has bias towards women's employment. Caliskan \emph{et al.} \cite{caliskan2017semantics} validated a gender bias through a word embedding model, which suggests that men's names have a stronger association with career related terms. In addition, language models may also generate racist biases \cite{wallace2019universal}, such as when using \emph{African American} as a prompt, the model may generate unpleasant text \cite{nadeem2020stereoset}.

\begin{figure*}[ht]
	\centering
	\includegraphics[width=0.8\textwidth]{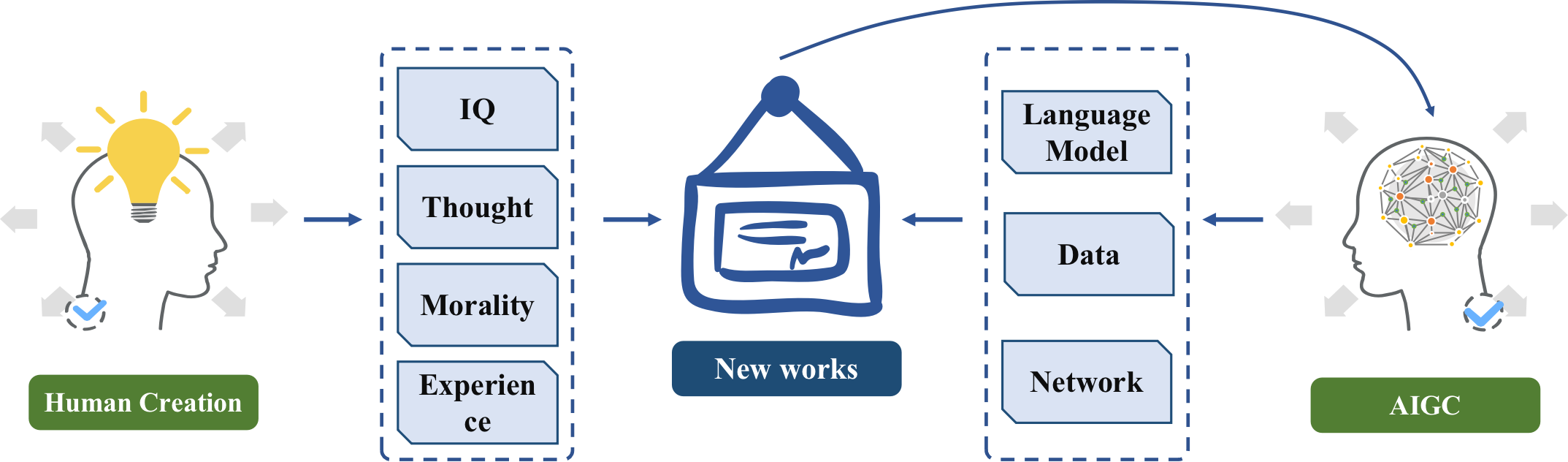}
	\caption{The similarity between human creations and AI generated content.}
	\label{fig: similarity}
\end{figure*}

\subsubsection{Camouflage}

The content generated by AI has camouflage, as shown in Fig. \ref{fig: similarity}. Through continuous iterative training with a large amount of data, AI models can produce content that is very similar to what humans create. This means that humans may not be able to identify synthetic content and real or human created content, which can cause a series of problems, including framing, malicious fraud, political manipulation, etc. \cite{paris2019deepfakes,lyu2020deepfake}. And with the development of generative AI models, this trend is becoming increasingly evident.

Deepfake is a typical representative of AIGC's camouflage, which is commonly used to generate fake images, speech, and videos, capable of synthesizing facial features from one person to another with astonishing accuracy, or creatively generating a video. The emergence of deepfakes has brought various threats to society. In the early days, deepfakes were used to exchange faces in videos, such as using celebrity faces in pornographic videos \cite{korshunov2018deepfakes,aliman2022vr}, creating celebrity scandals, affecting celebrities' reputations, and causing negative social impacts. Deepfakes can also play a role in advertising manipulation. Existing technologies can generate realistic advertisements, and even generate personalized advertisements for users based on personal data such as their height, race, and living environment \cite{campbell2022preparing}. These advertisements contain exaggerated and beautified content. The familiar stories in advertisements make users more inclined to believe in the authenticity of advertisements, which can affect consumers' purchasing decisions and infringe on their rights. With the development of technology, deepfakes can even be used to disrupt security systems. Due to the open-source nature of many deepfake technologies, criminals do not need to master a large amount of professional knowledge \cite{li2022seeing}, which poses a serious threat to social security. For example, a group of tax scammers infiltrated a Facial Liveness Verification system through open-source deepfake technology, causing significant economic losses \cite{borak2021chinese}.

Currently, there are also some detection methods for forged images and videos generated by deepfakes. However, these methods still have many loopholes. Hussain \emph{et al.} \cite{hussain2021adversarial} demonstrated that adversarial modification of synthesized fake videos can deceive existing detection methods to classify fake videos as real videos, and further demonstrated that this method is robust in both image and video fields. This indicates that the camouflage of AI generated content still poses a significant threat to the real world, and more technical methods need to be explored to distinguish AI generated content from the real world. DNN remains the most feasible method to address the threat posed by the camouflage of AI generated content \cite{gong2020deepfake}.

\subsubsection{Copyright}

The copyright ownership of AI generated works is a complex and challenging issue. Existing AI technology can generate exquisite art works, but whether these works should be protected by copyright has always been controversial \cite{sturm2019artificial,chen2023pathway}. In the past, work creation was once a human dominated skill and the concept of \emph{machine author} cannot be applied to copyright laws in most countries around the world \cite{he2019sentimental}. Due to the legal ambiguity of AI generated works, there have been some doubts and disputes over the copyright of AI generated works in recent years. As early as 2018, the artwork \emph{Portrait of Edmond Bellamy} created by AI was sold at a high price at an auction \cite{christie2018artificial}, raising questions about the copyright of AI generated works. More recently, Somepalli \emph{et al.} \cite{somepalli2022diffusion} demonstrated through evidence that the artistic works generated by the generative model Stable Diffusion copied its training data as output, infringing on the copyright of the creator.

The two core issues regarding the copyright of AI generated works are: a) whether the AI generated works have originality and whether they should be protected by copyright law, and b) if the copyright of AI generated works is recognized, then who should enjoy the copyright. In response to the first question, Cubert \emph{et al.} \cite{cubert2018law} argue that AI models may appear to be performing a complex task, but they are still programmed in nature and therefore lack creativity. In fact, the emergence of new technologies leading to copyright issues is not without precedent. In the era when cameras were first invented, American courts had dealt with similar issues and ultimately believed that the clothing, scenery, and other elements in the photographic work reflected the intention of the creator, thus recognizing the copyright of the photographic work \cite{he2019sentimental}, as the intention of the creator was considered an important characteristic of the original work \cite{mccormack2019autonomy}. AI generated works also typically require prompt input from users, and the extent to which these inputs can represent the creator's intentions remains an open question. The second issue is more complex because the completion of an AI generated work involves users, programmers, and human writers who provide prompt input, AI models, and training data, respectively. Unlike photographers, users who create works using AI still cannot know the final effect of the work after completing the input \cite{brown2018artificial}. This is because AI models perform complex processing on the input, and the final result inevitably includes certain features in the training data, which also involves infringement of human works \cite{cetinic2022understanding}.

To address these issues, one possible approach is to perform special processing on AI generated works, such as clearly marking the source and generation method of the generated works, or using special licensing or authorization mechanisms to manage the use and distribution of the generated works. In addition, it is necessary to modify the existing copyright legal framework to adapt to the unique nature of AI generated works. Overall, the issue of copyright ownership in AIGC generated works is complex and requires comprehensive consideration of technical, legal, and ethical factors, and currently remains a major challenge faced by AIGC.

\subsection{Threats from Malicious Users}

Not all AIGC users are well intentioned, and it must be acknowledged that the AIGC model itself may be attacked by malicious users, further threatening privacy and security in AIGC. This type of attack mainly includes model inversion attack, membership inference attack, poisoning attack, and model extraction attack.

\subsubsection{Model Inversion Attack}

Model inversion attacks extract training data from already trained models, with the aim of obtaining sensitive data behind the model, such as personal identity information and trade secrets. Once an attacker successfully extracts training data, they can abuse it, causing personal privacy breaches, copyright infringement, or other malicious behavior. In addition, attackers can also use the extracted training data to train their own models, resulting in further privacy leakage risks.

In AIGC, this attack method has been proven to be fully feasible \cite{behnia2022ew,tinsley2021face}. Taking a large language model as an example, attackers can accurately restore training data from the model \cite{chen2021knowledge}. Carlini \emph{et al.} \cite{carlini2021extracting} demonstrated that for large language models, individual training data can be extracted by querying the language model, and GPT-2 was used as the experimental object to extract personal information, including name, email, and phone number. Recently, the use of diffusion models for image generation has become increasingly popular. Carlini \emph{et al.} \cite{carlini2023extracting} conducted similar experiments on diffusion models. The author extracted over 1000 training samples from the state-of-the-art diffusion model, including personal photos, company logos, etc. And the experimental results show that the diffusion model has poorer privacy than the earlier generative model (such as GAN). Zhang \emph{et al.} \cite{zhang2020secret} theoretically analyzed and demonstrated that strong model capabilities are negatively correlated with their ability to resist model inversion attacks, and differential privacy has little effect on defending against such attacks, suggesting that AIGC models are more susceptible to model inversion attacks.

In the face of model inversion attacks, it is necessary to continuously improve and strengthen the security of the models. This includes using adversarial training, differential privacy, sensitive information filtering, and other means to enhance the robustness of the AIGC models, making it less vulnerable to model inversion attacks.

\subsubsection{Membership Inference Attack}

Membership inference attack is a privacy attack method aimed at inferring whether a specific sample is being used as training data for the model by observing the samples generated by the model or the output. Attackers can use these inference results to disclose user privacy information or steal sensitive data. Membership inference attack is a serious threat to AIGC. For example, in the medical field, where data sharing is difficult due to the sensitive nature of patient data, synthetic data is a potential solution. This method synthesizes data that is consistent with the original data distribution through AI but does not disclose true patient information \cite{zhang2022membership}. However, this method is easily attacked by membership inference. By inferring whether a patient's medical record is included in a model training set related to a certain disease, it is highly likely to believe that the patient has this specific disease \cite{hu2022membership}.

Currently, there have been many related researches on membership inference attacks against the generative model. Zhang \emph{et al.} \cite{zhang2022membership} conducted experiments on two datasets obtained from medical centers, and the research results showed that partially synthesized data is susceptible to membership inference attacks, while fully synthesized data is only slightly affected. This indicates that even if some of the original data is leaked, the overall harm is still significant. Chen \emph{et al.} \cite{chen2020gan} proposed a general attack model, which is suitable for various deep generative models. Experiments have verified that the method works well in three scenarios: image, medical data, and location data. Breugel \emph{et al.} \cite{van2023membership} proposed a density based membership inference attack method that is much more successful than previous work in attacking uncommon samples. Duan \emph{et al.} \cite{duan2023diffusion} proposed a black box membership inference attack method for diffusion models, assessing the matching of forward process posterior estimation at each timestamp. This method is suitable for both standard diffusion models such as DDPM and text-to-image diffusion models such as Stable Diffusion. Zhu \emph{et al.} \cite{zhu2023data} proposed new attack methods for each attack scenario specifically related to diffusion models, and achieved high attack performance (\textgreater0.9 AUCROC) in experiments. 

The membership inference attack methods against the generative model are constantly emerging, and show amazing attack performance. This fact shows that membership inference attack poses a serious challenge to AIGC security.

\subsubsection{Attacks Against Models}

The two attack methods mentioned earlier both target training data, which can lead to user privacy leakage, while the attack method against the model targets the model itself. Such attack methods have also been extensively studied. Although most of these studies focus on the discriminative model, in recent years, some attack methods against the generative model have also been proposed, suggesting that the AIGC model itself has security risks \cite{sun2021adversarial,guo2022threats}.

\begin{figure}[ht]
	\centering
	\includegraphics[width=0.47\textwidth]{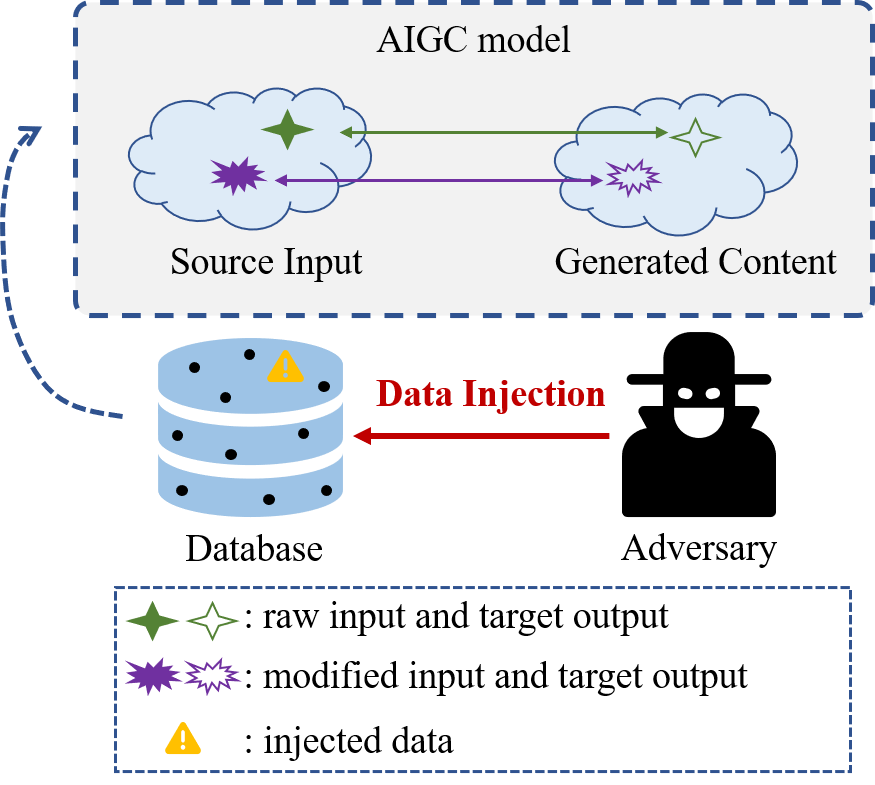}
	\caption{Poisoning attack against AIGC model.}
	\label{fig: poisoning}
\end{figure}

A common method of attacking models is a poisoning attack, as shown in Fig. \ref{fig: poisoning}. Poisoning attacks aim to intentionally create and inject malicious data samples, causing AIGC models to produce misleading, deceptive, or harmful results when generating content. Ding \emph{et al.} \cite{ding2019trojan} proved that the poisoning of training data, that is, the injection of customized data, can teach Trojan behavior to the deep generative model without affecting the original training target. Taking the autonomous driving scene as an example, the rain-removal model injected with poisoning data can modify the traffic lights with specific appearances (i.e. triggers) from red to green while removing raindrops. Such attacks may lead to the model misjudging the status of traffic lights, leading to misleading driving behavior and increasing the risk of accidents.

Another attack method is called model extraction attack, which aims to extract sensitive information from the model or reconstruct the internal structure of the model, which may lead to model abuse and pose a threat to the copyright of the model. Hu \emph{et al.} \cite{hu2021model} studied model extraction attacks against GANs from the perspective of accuracy extraction and fidelity extraction, and illustrated the feasibility of such theft through a case where a state-of-the-art GAN was stolen.

Although the attack method against the generative model has not been systematically studied yet, from the little work that has been done, this attack poses a serious threat to AIGC, which is still a challenge worthy of attention. Targeted enhancement of the robustness of AIGC models based on different attack methods, combined with digital watermarking, blockchain and other technologies to enhance copyright protection of AIGC models, is a possible solution to address these threats.

\subsection{Challenges in a Wider Range of Application Scenarios}

Due to its powerful capabilities, AIGC is increasingly being applied in practical scenarios. In addition to the unresolved challenges mentioned earlier, AIGC also faces some special problems in a wider range of scenarios. Below, we will explore AIGC edge deployment and privacy security issues faced by AIGC in the metaverse.

\subsubsection{Deploying AIGC in Edge Networks}

The centralized AIGC service has service latency, as mentioned in \cite{liu2023blockchain}, to generate images on the \emph{Hugging Face} platform, users must wait for about 1 minute. Firstly, this is because in a centralized AIGC deployment, the server needs to handle a large number of user requests, which require queuing after a user sends an image generation request. Secondly, the inference process of AIGC is also very time-consuming. The AIGC service deployed at the edge distributes service requests to servers closer to users, greatly reducing service latency and better supporting personalized services to improve user experience. Some researchers have also combined AIGC with the metaverse \cite{du2023generative,xu2023sparks} and proposed deploying AIGC services in edge networks to facilitate ubiquitous access by metaverse users.

However, edge deployed AIGC services can also pose additional privacy and security issues. In the edge deployed AIGC service, the data used for AIGC model training is stored in the edge device rather than concentrated on a central server. This decentralized approach increases the risk of data leakage \cite{xu2023unleashing}. Edge device may be threatened by physical security, network security, and vulnerabilities of the device itself, resulting in data being accessed or stolen by unauthorized personnel. In addition, due to the dispersion of data across multiple devices, more data transmission may increase the likelihood of data leakage. Moreover, edge device usually have limited security protection capabilities. Compared with centralized devices such as ECS, their ability to resist attacks is weaker. Attackers may take advantage of the weaknesses of edge device, such as vulnerabilities and unauthorized access to obtain AIGC models or related data, thereby endangering the privacy and security of users. To summarize, edge device have high physical accessibility, making it easier for attackers to carry out physical attacks or malicious tampering.

\subsubsection{AIGC in Metaverse}

The metaverse is a virtual digital world, which is a computer-generated virtual environment in which users can create, interact, and experience various virtual content. And AIGC can generate realistic multimedia content such as images, audio, and videos, which can provide rich and diverse virtual content for the metaverse, including virtual characters, virtual scenes, virtual objects, and so on. Through AIGC technology, it is possible to automatically generate virtual environments and create a more vivid and realistic metaverse experience for users. As mentioned in \cite{lee2023if}, AIGC can become an important technological driver of the metaverse, and there is a close connection between the two.

The frequent interaction between users and the metaverse fills the metaverse with sensitive data. In the context of the metaverse, AIGC faces more privacy and security issues \cite{chen2022metaverse}. AIGC technology is widely used in the metaverse to generate virtual characters, scenes, and interactive content. However, these generated content may involve the user's real personal information, such as facial features, sounds, behavioral habits, etc. \cite{sun2022metaverse}. If not authorized by the user or processed anonymously, these generated content may result in the leakage of the user's personal privacy. In addition, due to the fact that virtual identities in the metaverse are the digital representatives of users in that environment. Through AIGC technology, attackers may use the generative model to forge virtual identities, impersonate other users or create fake identities. This may lead to identity confusion, impersonation, identity theft, and other issues in the virtual world, posing a threat to user trust and security. Similarly, in the metaverse, virtual items and assets exist in digital form, and these generated items and assets may be subject to theft, forgery, tampering, or illegal copying, posing a threat to the security of users' property. The privacy and security issues existing in AIGC have been further amplified in the metaverse. In order for AIGC to truly become a key technology driving the development of the metaverse, these privacy and security issues must be taken seriously and resolved.

\section{Countermeasures}
It is clear from the description in Section III that generative AI is rapidly evolving and becoming more and more connected to human society. This trend of development also exposes the privacy and security issues of AIGC. Therefore, how to apply existing privacy computing and various security techniques to generative AI becomes a topic that has to be faced today. Privacy and security of data is one of the keys to make AIGC serves human society better. Only by properly handling the privacy and security issues brought by AIGC can we promote the sustainable development of generative AI. In the following, we will describe the existing techniques for privacy protection and ensuring data security in generative AI.

\begin{figure*}[ht]
	\centering
	\includegraphics[width=0.8\textwidth]{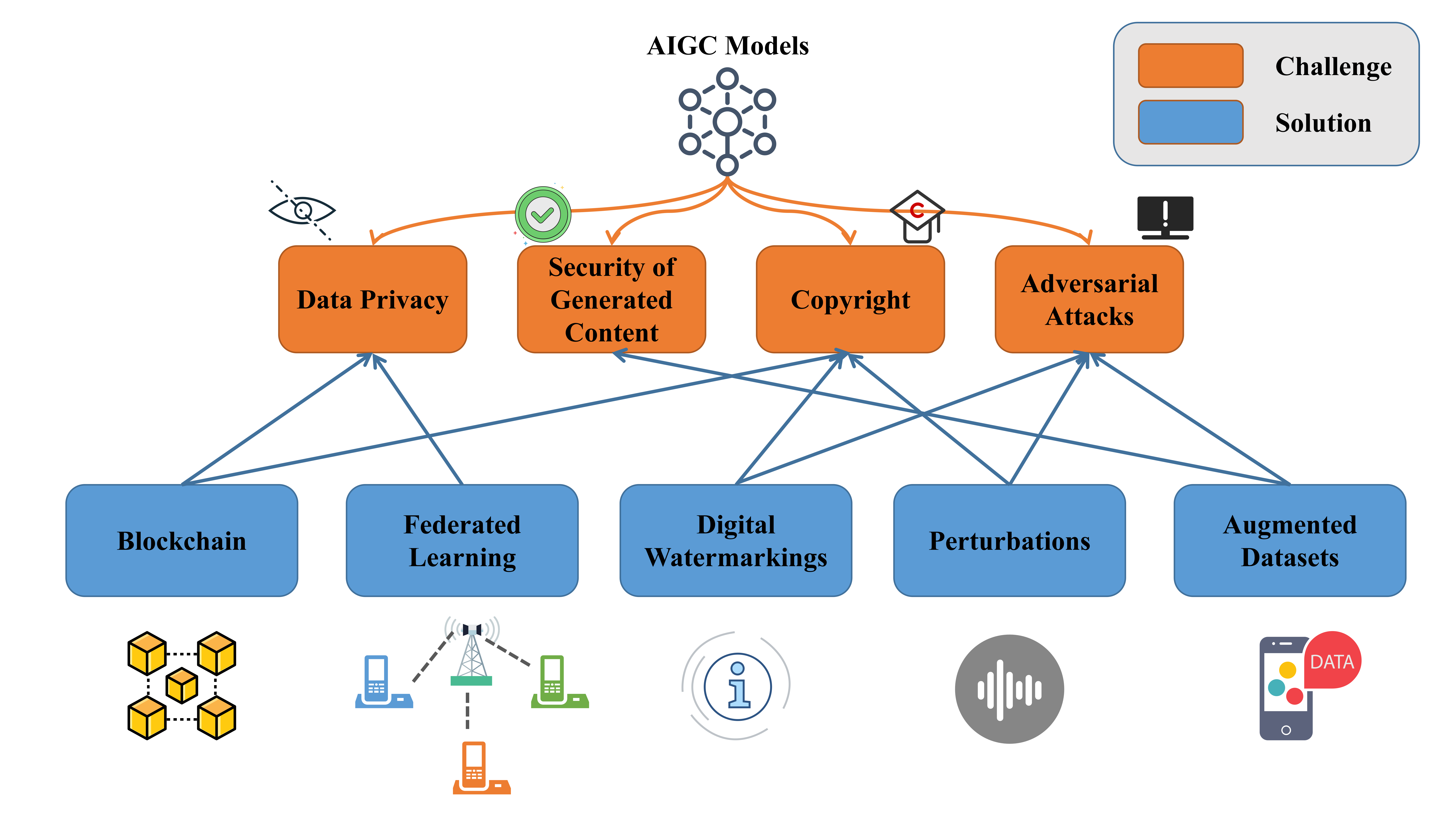}\hspace{-1pt}
	\caption{Relevant solutions to address the privacy and security challenges of AIGC which are summarized in this paper.}
	\label{fig: solutions}
\end{figure*}

\subsection{Data Privacy}

It is well known that large base models are vulnerable to privacy risks, and AIGC models built based on these models may also be vulnerable to privacy breaches. Companies and researchers try to protect user data privacy in a variety of ways. In addition to direct approaches such as providing warning messages to users, tools such as privacy computing and blockchain have become important supports for privacy protection in AIGC.

\subsubsection{Detecting Replicated Content}One straightforward way to protect the privacy of AIGC tool users' data is to detect duplicate content and send a warning message to the user \cite{chen2023pathway}. Stability AI points out that Stable Diffusion has limitations. For example, it has not yet been able to detect replicated images in the training data. For this purpose they provide a website \cite{beaumont2022clip} to identify such memorized images. In addition, the art company Spawning AI has created a website called "Have I Been Trained" \cite{spawning2022have} to help users determine whether their photos or artwork are being used as data for AI training. OpenAI proposes another measure to reduce data duplication through deduplication for the purpose of addressing data privacy issues in AIGC \cite{Nichol2022dalle}. Somepalli \emph{et al.} studied image retrieval frameworks to identify duplicated content \cite{somepalli2022diffusion}.

\subsubsection{Federated Learning}
During the lifecycle of the AIGC service, the large-scale dataset used for training and the user's private information need to be kept secure to avoid privacy breaches. Since AIGC applications are highly integrated with the Internet, the generation and storage of data used for AIGC model training occur on edge servers and mobile devices \cite{li2021federated}. As opposed to centralized data centers, edge and mobile devices are less capable of defending against a variety of privacy-threatening attacks. Recently, there are several distributed learning frameworks for privacy protection. Federated learning \cite{lim2020federated} is used to perform model fine-tuning and inference on mobile AIGC networks that meet privacy-preserving requirements. Federated learning, as a distributed machine learning method, does not transmit raw data during training, but local models, which can provide privacy and security for the operation of the AIGC network \cite{kang2022communication}. There are two specific ways that Federated Learning is used in the AIGC model.

\begin{itemize}
    \item Differential privacy: Differential privacy prevents the central server from identifying the owner of local updates.
    \item Secure aggregation: During FL process, individual mobile devices send local updates to the central server for global model aggregation. The aggregation process uses authentication, secret sharing mechanisms, etc. to ensure security.
\end{itemize}

\begin{figure*}
	\centering
	\includegraphics[width=0.8\textwidth]{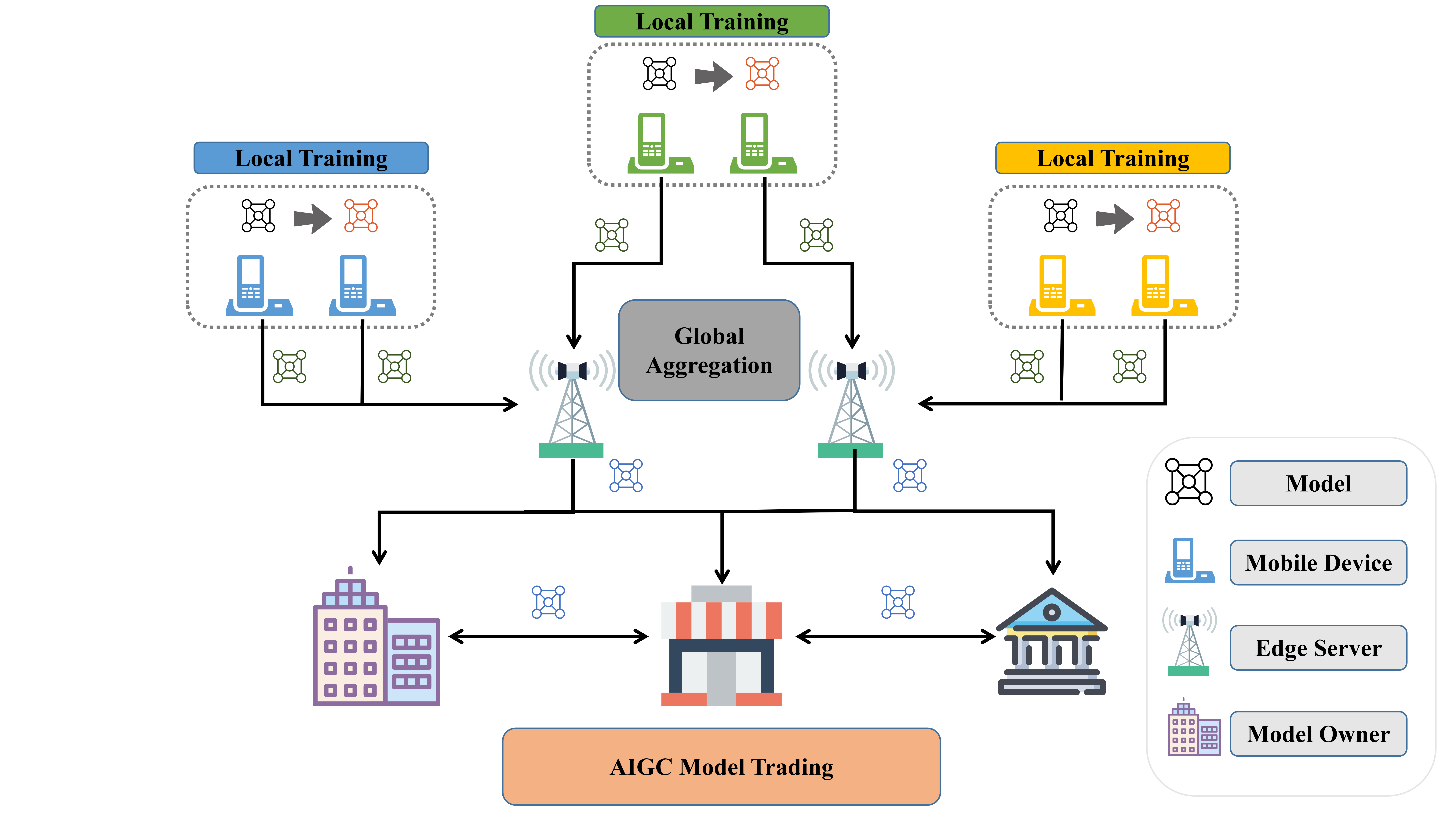}\hspace{-1pt}
	\caption{Federated Learning in AIGC mobile network. The local models are trained at mobile devices. The global models are aggregated at edge servers. The AIGC models are traded among their owners.}
	\label{fig: FLinAIGC}
\end{figure*}

Augenstein \emph{et al.} \cite{augenstein2019generative} proposed a differential private
federated generative model to synthesize representative examples of private data. The model can solve many data problems without human intervention and with guaranteed privacy. Fan \emph{et al.} \cite{fan2020federated} proposed an FL-based generative learning scheme to improve the efficiency and robustness of GAN models. This scheme is particularly effective in environments with highly skewed data distribution. Chung \emph{et al.} \cite{chung2022federated} proposed an unsupervised Iterative Federated Clustering algorithm. The algorithm uses generative models to deal with the statistical heterogeneity that may exist among the participants of FL, in order to find an inherent cluster structure in users’ data and unlabeled datasets. However, the centralized FL framework in \cite{fan2020federated}, \cite{chung2022federated} can lead to the risk of a single point-failure. Therefore, decentralized federation learning schemes can be considered to better protect data privacy in the AIGC mobile network. Che \emph{et al.} \cite{che2022decentralized} proposed a novel serverless federated learning framework Committee Mechanism based Federated Learning (CMFL), which can ensure the robustness of the algorithm with convergence guarantee. Wang \emph{et al.} \cite{wang2021efficient} proposed a decentralized FL framework based on a ring topology and deeply generated models. It contains a method for synchronizing the ring topology which can improve the communication efficiency and reliability of the system. Moreover, generative models can solve data-related problems, such as incompleteness, low quality,
insufficient quantity, and sensitivity. Additionally, Dockhorn \emph{et al.} \cite{dockhorn2022differentially} proposed a differentially private diffusion model to guarantee privacy in generative models.

\subsubsection{Blockchain}

In order to integrate AIGC with the Internet, people started to use mobile devices and edge devices to implement AIGC services. For example, EdgeMatrix \cite{shen2022edgematrix} is a multi-tier edge-cloud computing framework which is able to maximize the throughput of the system. Such a framework provides a way for mobile AIGC services to be implemented. In this case, a trustworthy collaborative AIGC service provisioning framework must be established to protect the privacy of data in circulation.

On the other hand, mobile and edge AIGC providers can customize AIGC services by collaborating with many user nodes while distributing data to different devices \cite{lin2022novel}. As a result, the joint participation of multiple parties' content requires a secure access control mechanism to ensure privacy and security.

To address both of these challenges, the use of blockchain technology comes naturally to mind. Blockchain-based on distributed ledger technology can be used to explore a secure and reliable AIGC service provisioning framework, record resource and service transactions, encourage data sharing among nodes, and form a trustworthy AIGC ecosystem.

\begin{figure*}[ht]
	\centering
	\includegraphics[width=0.8\textwidth]{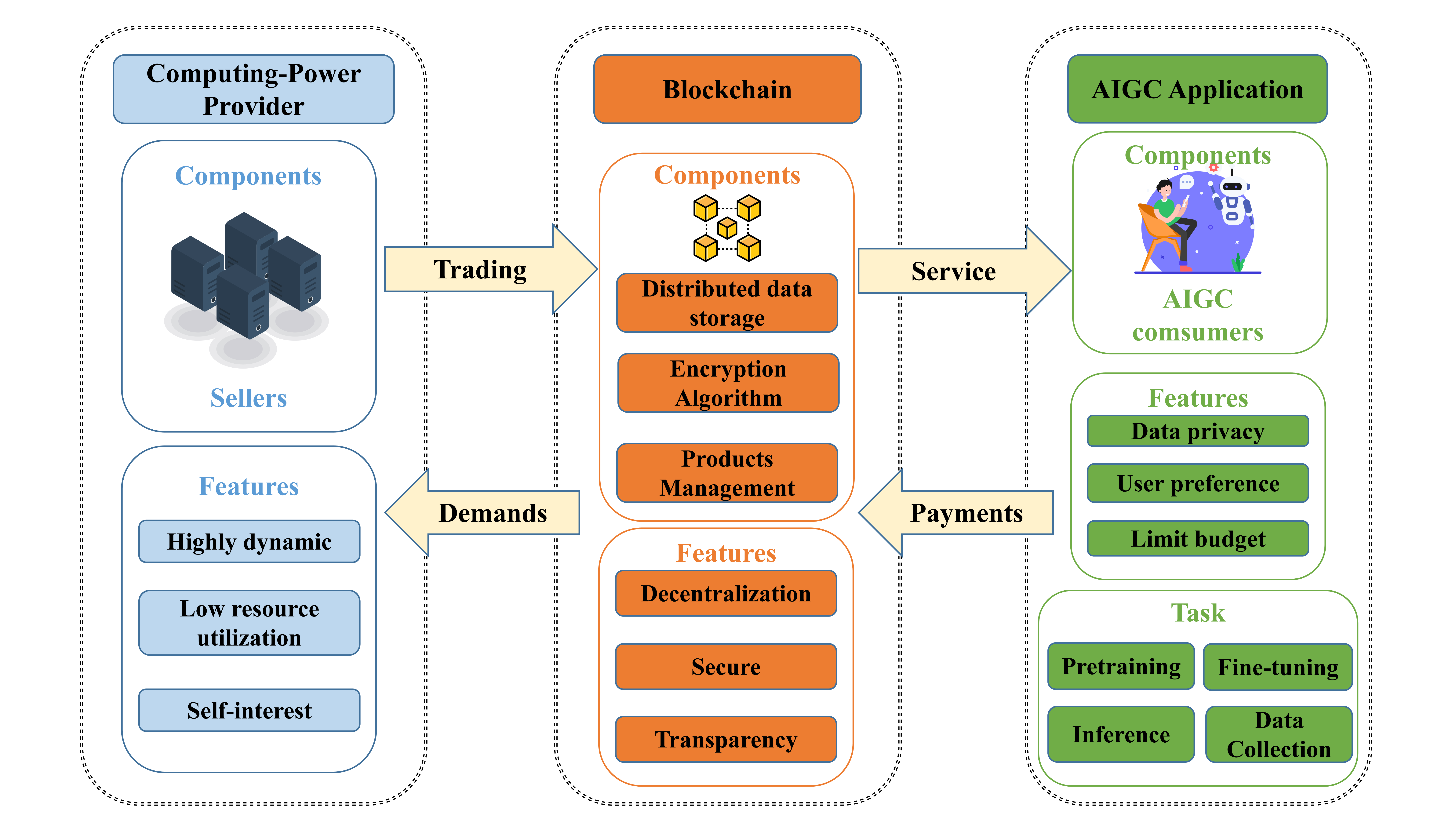}\hspace{-1pt}
	\caption{Blockchain in AIGC mobile network \cite{ren2022ai}. In order to provide AIGC services, the framework includes computing-power provider, blockchain and AIGC application. }
	\label{fig: BVinAIGC}
\end{figure*}

Mobile AIGC networks rely on cloud computing technology, and fusing blockchain technology with existing cloud systems has a great potential in both functionality/performance enhancement and security/privacy improvement \cite{gai2020blockchain}. Lin \emph{et al.} \cite{lin2023unified} proposed a unified blockchain-semantic ecosystems framework for wireless edge intelligence-enabled Web 3.0. It is able to analyze the semantic information of contents that can convey precisely the desired meanings without consuming many resources, which provides a more reliable guarantee for the AIGC network. To further unleash the advantages of semantic extraction and communication in Web 3.0, a blockchain-based semantic exchange framework is proposed in \cite{lin2022blockchain} to realize fair and efficient interactions. Xu \emph{et al.} \cite{xu2022quantum} introduce a quantum blockchain-driven Web 3.0 framework that provides information-theoretic security for decentralized data transferring and payment transactions. Ren \emph{et al.} \cite{ren2022ai} proposed a computing-power trading framework based on blockchain, also named AI-Bazzar. In AI-Bazzar, the AI consumer plays multiple roles and feels free to contribute the computing power rented from the
computing-power provider (CPP) for blockchain mining and AI services. This framework is also applicable to the deployment of AIGC services in mobile networks.

Blockchain-based approaches have also been linked to generative AI tools that play a role for data privacy and security. Dirgantoro \emph{et al.} \cite{dirgantoro2020generative} propose an edge intelligence framework based on deep generative models and blockchain. To overcome the accuracy issue of the limited dataset, GAN is leveraged in the framework to synthesize training samples. Then, the output of this framework is confirmed and incentivized by smart contracts based on the proof-of-work consensus algorithm.

The blockchain mechanism can also be combined with the aforementioned federal learning to jointly maintain the privacy of user data in AIGC.  Li \emph{et al.} \cite{li2020blockchain} proposed a decentralized federated learning framework based on blockchain, that is, a Blockchain-based Federated Learning framework with Committee consensus (BFLC). Without a centralized server, the framework uses blockchain for the global model storage and the local model update exchange. To enable the proposed BFLC, they also devised an innovative committee consensus mechanism, which can effectively reduce the amountof consensus computing and reduce malicious attacks. This further enhances the protection of users' privacy when Federal Learning is combined with AIGC. 

What's more, the multimodal outputs of AIGC can be minted as NFTs and then recorded on the blockchain. Tann \emph{et al.} \cite{tann2022predicting} develop a conditional generative model to synthesize new digital asset collections based on the historical transaction results of previous collections. 

As summarized in \cite{xu2023unleashing}, there are several benefits that blockchain brings to mobile AIGC networks:

\begin{itemize}
    \item Data Administration: The use of blockchain with smart contracts to record transactions of resources and services in the AIGC network makes data management in AIGC credible.
    \item Computing and Communication Management: Blockchain allows heterogeneous computing resources to be managed securely and efficiently in the mobile AIGC network.
    \item Optimization: The blockchain always provides available and complete historical data that can be used to optimize the services of the AIGC network.
\end{itemize}

\subsection{Security of Generated Content}

The most straightforward way to secure AIGC applications is to evaluate and control these generative contents. Current solutions work on three main perspectives of AIGC: factuality, toxicity and identifiability.

\subsubsection{Factuality}

Although AI tools like ChatGPT can generate content that seems to make sense, they are often not reliable enough when it comes to factuality. If the model outputs absurd or even completely opposite conclusions to the facts, it will pose a great threat to the authenticity of the information on the Internet.

The first step in regulating AIGC is to find indicators that can reasonably assess the factuality of the content. Some early research showed that AI models can produce results that are contrary to the facts \cite{zhang2018texttruth}. To better estimate the truthfulness of the model output, Goodrich \emph{et al.} \cite{goodrich2019assessing} propose a model-based metric to evaluate the factual accuracy of generated text that is complementary to typical scoring schemes like ROUGE (Recall-Oriented Understudy for Gisting Evaluation) \cite{lin2004rouge} and BLEU (Bilingual Evaluation Understudy) \cite{papineni2002bleu}. With the rise of generative AI, Truthful AI \cite{evans2021truthful} proposes systematic definitions of truthfulness standards and approaches for governing AIGC. Alaa \emph{et al.} \cite{alaa2022faithful} introduced a 3-dimensional evaluation metric that characterizes the fidelity, diversity and generalization performance of any generative model in a domain-agnostic fashion.

For language models, a variety of work has been done to constrain the generated language content in different ways. Based on GPT-3, WebGPT \cite{nakano2021webgpt}  proposes a humanoid prototype that models the AI response process as a web searching and evidence composing phrases. This has resulted in a significant improvement in the factual accuracy of AIGC on several benchmark datasets \cite{lin2021truthfulqa}. To measure and improve the factual accuracy of large language models used for text generation, Lee \emph{et al.} \cite{lee2022factuality} proposes the factual-nucleus sampling algorithms that strike a balance between factuality and quality of AIGC. Azaria \emph{et al.} \cite{azaria2023internal} introduces a simple yet effective method to detect the truthfulness of LLM-generated statements, which utilizes the LLM's hidden layer activations to determine the veracity of statements. A classifier is trained to detect which statement is true or false based on an LLM's activation values.

\subsubsection{Toxicity}

In addition to the factuality of the content, the fairness and harmlessness of the language model output is also important. If AI tools generate content that is biased and inconsistent with mainstream human values, it will have a negative impact on human society.

OpenAI proposes InstructGPT \cite{ouyang2022training}, which shows an avenue for aligning language models with user intent on a wide range of tasks by fine-tuning with human feedback. Google proposes LaMDA \cite{thoppilan2022lamda}, a family of Transformer-based neural language models specialized for dialog. It demonstrates that fine-tuning with annotated data and enabling the model to consult external knowledge sources can lead to significant improvements towards the two key challenges of safety and factual grounding. Furthermore, Ganguli \emph{et al.} \cite{ganguli2022red} studies and improve the safety of language models in an adversarial way. They investigate scaling behaviors for red teaming across 3 model sizes (2.7B, 13B, and 52B parameters) and 4 model types. They find that the RLHF (Reinforcement Learning from Human Feedback) models are increasingly difficult to red team as they scale. Sison \emph{et al.} \cite{sison2023chatgpt} propose the Human-Centered Artificial Intelligence (HCAI) framework. It provides objectives, principles, procedures, and structures for reliable, safe, and trustworthy AI. To better identify biases and injustices in AIGC, it is essential to audit trained generative networks in a human-understandable form. Olson \emph{et al.} \cite{olson2023cross} proposed Cross-GAN Auditing (xGA), an approach that compares a newly developed GAN against a prior baseline. This provides users and model developers with an intuitive assessment of the similarities and differences between GANs, which in turn provides ideas for measuring imbalances and biases in AIGC models.

\subsubsection{Identifiability} 

The flourishing of AIGC has brought more options for human beings to work, study and create. However, as the characteristics of AIGC become closer and closer to human-created content, it is increasingly difficult to tell if certain works are generated by AI. For this reason, there has been a significant amount of work trying to distinguish them from each other. How to improve the identifiability of AIGC is also an important part to ensure the safety of generated content.

The rapid improvement of language models has raised the specter of abuse of text generation systems. GLTR \cite{gehrmann2019gltr} is a tool to support humans in detecting whether a text was generated by a model. It applies a suite of baseline statistical methods that can detect generation artifacts across common sampling schemes. A feature-based detection method of automated language models is proposed in \cite{frohling2021feature} . It uses carefully crafted features that attempt to model intrinsic differences between human and machine text. Guo \emph{et al.} \cite{guo2023close} conducted comprehensive human evaluations and linguistic analyses of ChatGPT-generated content compared with that of humans, where many interesting results are revealed. Kumarage \emph{et al.} \cite{kumarage2023stylometric} present a novel algorithm using stylometric signals to aid in detecting AI-generated tweets. Their extensive experiments demonstrate that the stylometric features are effective in augmenting the state-of-the-art AI-generated text detectors. Yu \emph{et al.} \cite{yu2023cheat} initially investigate the possible negative impact of ChatGPT on academia, and present a large-scale CHatGPT-writtEn AbsTract (CHEAT) to support the development of detection algorithms. Liu \emph{et al.} \cite{liu2023argugpt} test existing AIGC detectors and build a detector using SVMs and RoBERTa. Results suggest that a RoBERTa fine-tuned with the training set of ArguGPT achieves above 90\% accuracy in both essay- and sentence-level classification. Kirchenbauer \emph{et al.} \cite{kirchenbauer2023watermark} proposed a watermark to detect whether the text is generated by AI models. They have only tested it on Meta's smaller open source language model, OPT-6.7B, so its effectiveness on larger, widely used language models like ChatGPT is unknown. Overall, these detection methods still have some drawbacks at present. Liang \emph{et al.} \cite{liang2023gpt} evaluate the performance of several widely-used GPT detectors using writing samples from native and non-native English writers. Their findings reveal that these detectors consistently misclassify non-native English writing samples as AI-generated, whereas native writing samples are accurately identified.

Additionally, the identifiability of AI-generated images is receiving increasing attention. Lu \emph{et al.} \cite{lu2023seeing} conducted a high-quality quantitative study with fifty participants. They observe that there tend to be certain defects in AI-generated images that
serve as cues for people to distinguish between real and fake photos. For example, AI-generated images may lack fine details, such as wrinkles in clothing or hair details, and they may appear smoother or more uniform than real photos, such as
smooth skin or unrealistic facial expressions. These defects provide ideas on how to discriminate AI-generated images in the future.

\begin{table*}[h]
\centering
\renewcommand{\arraystretch}{1.3}
\caption{Summary of methods to ensure the security of generated content}
\resizebox{\columnwidth}{!}{
\begin{tabular}{|l|l|l|l|}
\hline
\textbf{Scenarios}                        & \textbf{Tools}                                                                                             & \textbf{References} & \textbf{Brief Description}                                                                                                                                                 \\ \hline
\multirow{6}{*}{Factuality}      & Model-based metric                                                                                &    \cite{goodrich2019assessing}        & \begin{tabular}[c]{@{}l@{}}A model-based metric to evaluate the factual accuracy \\ of generated text\end{tabular}                                                \\ \cline{2-4} 
                                 & Truthful AI                                                                                       &      \cite{evans2021truthful}      & \begin{tabular}[c]{@{}l@{}}Systematic definitions of truthfulness standards and \\ approaches for governing AIGC\end{tabular}                                     \\ \cline{2-4} 
                                 & 3-dimensional evaluation metric                                                                                       &      \cite{alaa2022faithful}      & \begin{tabular}[c]{@{}l@{}}An evaluation that characterizes the fidelity, diversity \\ and generalization performance of any generative model \end{tabular}                                     \\ \cline{2-4}
                                 & WebGPT                                                                                            &    \cite{nakano2021webgpt}        & \begin{tabular}[c]{@{}l@{}}A humanoid prototype that models the AI response process \\ as a web searching and evidence composing phrases\end{tabular}             \\ \cline{2-4} 
                                 & Factual-nucleus sampling algorithms                                                               &  \cite{lee2022factuality}          & \begin{tabular}[c]{@{}l@{}}Dynamically adapts the randomness to improve the \\ factuality of generation\end{tabular}                                              \\ \cline{2-4} 
                                 & LLM's hidden layer activations                                                                    &     \cite{azaria2023internal}       & \begin{tabular}[c]{@{}l@{}}A method which utilizes the LLM's hidden layer \\ activations to determine the veracity of statements\end{tabular}                     \\ \hline
\multirow{5}{*}{Toxicity}        & InstructGPT                                                                                       &     \cite{ouyang2022training}       & Fine tunes the language models with human feedback                                                                                                                \\ \cline{2-4} 
                                 & LaMDA                                                                                             &    \cite{thoppilan2022lamda}        & \begin{tabular}[c]{@{}l@{}}Fine tunes with annotated data and enables the model \\ to consult external knowledge sources\end{tabular}                             \\ \cline{2-4} 
                                 & Red teaming language models                                                                       &   \cite{ganguli2022red}         & \begin{tabular}[c]{@{}l@{}}Improves the safety of language models in an adversarial \\ way\end{tabular}                                                           \\ \cline{2-4} 
                                 & \begin{tabular}[c]{@{}l@{}}Human-Centered Artificial Intelligence\\ (HCAI) framework\end{tabular} &    \cite{sison2023chatgpt}        & \begin{tabular}[c]{@{}l@{}}Provides objectives, principles, procedures and \\ structures for reliable, safe, and trustworthy AI\end{tabular}                      \\ \cline{2-4}
                                 & xGA                                                                       &   \cite{olson2023cross}         & \begin{tabular}[c]{@{}l@{}}Compares a newly developed GAN against a prior baseline\end{tabular}                                                           \\ \hline
\multirow{7}{*}{Identifiability} & GLTR                                                                                              &      \cite{gehrmann2019gltr}      & \begin{tabular}[c]{@{}l@{}}Applies a suite of baseline statistical methods that \\ can detect generation artifacts across common sampling \\ schemes\end{tabular} \\ \cline{2-4} 
                                 & Feature-based detection                                                                           &    \cite{frohling2021feature}        & \begin{tabular}[c]{@{}l@{}}A simple feature-based classifier for detecting machine \\ generations among human-written text\end{tabular}                           \\ \cline{2-4} 
                                 & Human ChatGPT Comparison Corpus (HC3)                                                             &    \cite{guo2023close}        & \begin{tabular}[c]{@{}l@{}}A dataset which includes comparison responses from both \\ human experts and ChatGPT\end{tabular}                                      \\ \cline{2-4} 
                                 & Stylometric signals                                                                               &   \cite{kumarage2023stylometric}         & \begin{tabular}[c]{@{}l@{}}A novel algorithm using stylometric signals to aid \\ detecting AI-generated tweet\end{tabular}                                        \\ \cline{2-4} 
                                 & \begin{tabular}[c]{@{}l@{}}CHatGPT-writtEn AbsTract dataset\\ (CHEAT)\end{tabular}                &     \cite{yu2023cheat}       & \begin{tabular}[c]{@{}l@{}}A large-scale dataset to support the development of \\ detection algorithms\end{tabular}                                               \\ \cline{2-4} 
                                 & ArguGPT                                                                                           &       \cite{liu2023argugpt}     & \begin{tabular}[c]{@{}l@{}}A balanced corpus of argumentative essays generated \\ by several GPT models\end{tabular}                                              \\ \cline{2-4} 
                                 & Watermark                                                                                         &     \cite{kirchenbauer2023watermark}       & \begin{tabular}[c]{@{}l@{}}A watermark for detecting whether the text is \\ generated by AI models\end{tabular}                                                    \\ \hline
\end{tabular}}
\end{table*}

\subsection{Copyright}

Various AIGC applications have brought more choices of creative tools to people. Meanwhile, this also makes the copyright protection of AIGC a new issue that needs to be solved urgently. The root cause of the high concern about intellectual property rights in AIGC, especially copyright infringement, is that the formation and refinement of AIGC models rely on a large amount of data training, and the data used for training often contains content protected by copyright law. We summarize the technologies that can be used for AIGC copyright protection from the following aspects: blockchain, digital watermarking, adversarial perturbation, and attribution.

\subsubsection{Laws and Regulations}To deal with copyright issues, many AIGC companies have established some regulations. Midjourney, for example, has added a DMCA (Digital Millennium Copyright Act) removal policy to its terms of service. Artists can request that their work be removed from the dataset if they believe Midjourney is infringing on their copyrights \cite{midjourney2022term}. Stability AI has a similar approach, and they plan to give artists the option of excluding themselves from future versions of Stable Diffusion \cite{Heikkila2022artists}.

\subsubsection{Blockchain}Blockchain has become an important tool for AIGC's copyright protection. It has been used for a long time in the field of rights confirmation. Yang \emph{et al.} \cite{yang2022trustworthy} proposed a decentralized blockchain-based FL (B-FL) architecture by using a secure global aggregation algorithm to resist malicious devices, and deploying a practical Byzantine fault tolerance consensus protocol with high effectiveness and low energy consumption among multiple edge servers to prevent model tampering from the malicious server. This is an example of parties building trust in the network with the help of blockchain, and this trust can also be introduced into the AIGC network for the task of confirming rights.

To guarantee the ownership of digital objects, economic systems based on public blockchains and Non-Fungible Tokens (NFTs) are beginning to emerge. In order to combat NFT plagiarism, Garcia \emph{et al.} \cite{garcia2022semantics} proposed CopyrightLY, a decentralized application for authorship and copyright management. Content and metadata are stored in decentralized storage and registered on the blockchain. This tool that uses NFT to store data and protect ownership can be used in emerging scenarios such as metaverse, and likewise provides ideas for AIGC copyright protection.

Liu \emph{et al.} \cite{liu2023blockchain} propose a protocol to protect the ownership and copyright of AIGC, called Proof-of-AIGC. The Proof-of-AIGC consists of
two phases, namely proof generation and challenge. Proof generation registers
AIGC products on blockchain. Given the immutability of blockchain ledger, the concerns about ownership tampering can be effectively addressed. If the producer of AIGC finds another product that is significantly similar to its own work, it can initialize the challenge process by sending a transaction. The Proof-of-AIGC solves the ownership tampering and plagiarization of AIGC products.

\subsubsection{Digital Watermarking}

A digital watermark \cite{cox2007digital} is a kind of marker covertly embedded in a noise-tolerant signal such as audio, video or image data. It is typically used to identify ownership of the copyright of such signal.

Deep learning and data hiding have been combined early to drive advances in secure communications and authentication \cite{byrnes2021data}. Some deep watermarking models adopt the encoder-decoder framework without including a discriminator. The potential of the CNN-based encoder-decoder frameworks for digital image watermarking was first explored in \cite{kandi2017exploring}, which uses two traditional Convolutional Auto-Encoders for watermark embedding and extraction. WMNet \cite{mun2019finding} uses a relatively shallow network to achieve robust and blind digital watermarking results. Zhong \emph{et al.} \cite{zhong2020automated} generalize the watermarking process by training a deep neural network to learn the general rules of watermark embedding and extraction so that it can be used for a range of applications and combat unexpected distortions. ReDMark \cite{ahmadi2020redmark} uses two Full Convolutional Neural Networks (FCNs) for embedding and extraction along with a
differentiable attack layer to simulate different distortions,
creating an end-to-end training scheme.

Another primary approach for deep watermarking uses GANs \cite{goodfellow2014generative}. The first end-to-end trainable framework for data hiding was a model called HiDDeN \cite{zhu2018hidden}, which uses an adversarial discriminator to improve performance. ROMark \cite{wen2019romark} builds upon the framework from HiDDeN \cite{zhu2018hidden} by using a min-max formulation for robust optimization. A novel embedding strategy using Inverse Gradient Attention (IGA) \cite{zhang2020robust} was adopted. It uses IGA mask to improve capacity and robustness. Liu \emph{et al.} \cite{liu2019novel} introduced a novel two-stage separable deep learning (TSDL) framework for watermarking. Some methods adopted the Wasserstein GAN (WGAN) \cite{arjovsky2017wasserstein} framework for their watermarking model. For example, Zhang \emph{et al.} \cite{zhang2019steganogan} introduced SteganoGAN. Using adversarial training, this model achieves a relatively high payload capacity.

With the rise of generative AI tools, the digital watermarking techniques described above can be used for AIGC copyright protection. Yadollahi \emph{et al.} \cite{yadollahi2021robust} proposed a framework for watermarking a DNN model designed for textual domain. The watermark generation scheme provides a secure watermarking method by combining Term Frequency (TF) and Inverse Document Frequency (IDF) of a particular word. Fei \emph{et al.} \cite{fei2022supervised} proposed a watermarking method for protecting the Intellectual Property (IP) of Generative Adversarial Networks (GANs). The aim is to watermark the GAN model so that any image generated by the GAN contains an invisible watermark (signature), whose presence inside the image can be checked at a later stage for ownership verification. Huang \emph{et al.} \cite{huang2022cmua} proposed a method that can generate a Cross-Model Universal Adversarial Watermark (CMUA-Watermark), protecting a large number of facial images from multiple deepfake models. It can effectively distort the fake facial images generated by multiple deepfake models. Fan \emph{et al.} \cite{fan2022pcpt} proposed two copyright protection and traceability framework. One is a Passive Copyright Protection and Traceability framework (PCPT), which is based on the idea of black-box neural network watermarking with the video framing and image perceptual hash algorithm. Another is a DNN model Active Copyright Protection and Traceability framework (ACPT), which is based on the authorization control strategy and image perceptual hash algorithm. They improve the existing traceability mechanism and enhance the framework security. Lim \emph{et al.} \cite{lim2022protect} demonstrated that the current digital watermarking framework is insufficient to protect image captioning tasks that are often regarded as one of the frontiers AI problems. They studied and proposed two different embedding schemes in the hidden memory state of a recurrent neural network to protect the image captioning model.

In addition, the text watermarking \cite{he2022protecting} previously used to protect the language generation API can be used to identify if the AIGC tool is using unauthorized store samples. For example, Stable Diffusion generates images that contain the watermark of Getty Images, which can pose serious copyright issues \cite{Vincent2023getty}.

\subsubsection{Perturbations}

Early studies \cite{szegedy2013intriguing} found that the input-output mapping for deep neural network learning is largely discontinuous. We can make DNNs output wrong results by applying a certain imperceptible perturbation. Unlike using perturbations for adversarial attacks, some studies \cite{salman2021unadversarial} \cite{wang2022defensive} use them to assist DNNs in decision making. Inspired by this, some works have investigated the feasibility of exploiting adversarial perturbations in the AIGC copyright protection scenario.

Xue \emph{et al.} \cite{xue2022advparams} encrypt a small number of model parameters by perturbing them with well-crafted adversarial perturbations. With the encrypted parameters, the accuracy of the DNN model drops significantly, which can prevent malicious infringers from using the model. After the encryption, the positions of encrypted parameters and the values of the added adversarial perturbations form a secret key. Authorized user can use the secret key to decrypt the model on machine learning as a service, while unauthorized user cannot use the model.

\begin{figure*}[ht]
	\centering
	\includegraphics[width=1\textwidth]{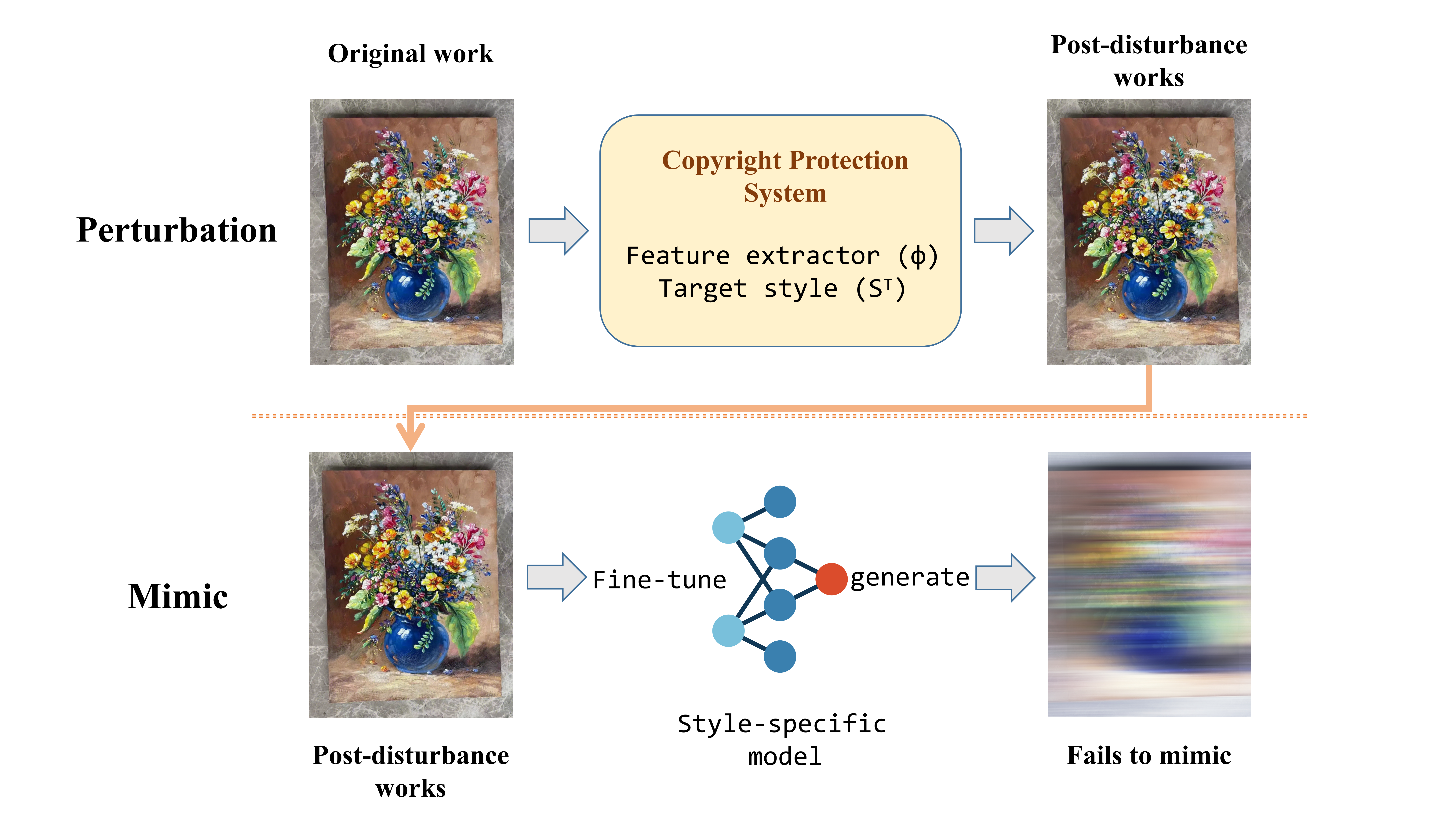}
	\caption{An overview of a method to prevent works from being mimicked by AIGC models through perturbation. The copyright protection system applies a tiny perturbation to the artist's original work which is not visible to the human eye. But when the AIGC model tries to mimic the style of the work after this perturbation, it will be misled and the generated work belongs to the target style rather than the original style. An example of this type of method is \cite{shan2023glaze}.}
	\label{fig: perturbation}
\end{figure*}

After using a perturbation-based approach to encrypt DNNs, the approach was explored for the copyright protection of AIGCs. Some people have taken the open sourced model, and “fine-tuned” it on additional samples from specific artists \cite{heikkilaarchive2022artist}. This is a serious threat to the copyright of artists and their works. To deal with this problem, Shan \emph{et al.} \cite{shan2023glaze} proposed \emph{Glaze}, a tool that enables artists to apply "style cloaks" to their art before sharing online. These cloaks apply barely perceptible perturbations to images, and when used as training data, mislead generative models that try to mimic a specific artist. Wang \emph{et al.} \cite{wang2023plug} proposed a plug-and-play invisible copyright protection method based on defensive perturbation
for DNN-based applications. They project the copyright information to the defensive perturbation with the designed copyright encoder, which is added to the image to be protected. Then, the method extracts the copyright information from the encoded copyrighted image with the devised copyright decoder. It correctly extracts the copyright information in the encoded image saved from social media.

\begin{figure*}[ht]
	\centering
	\includegraphics[width=0.8\textwidth]{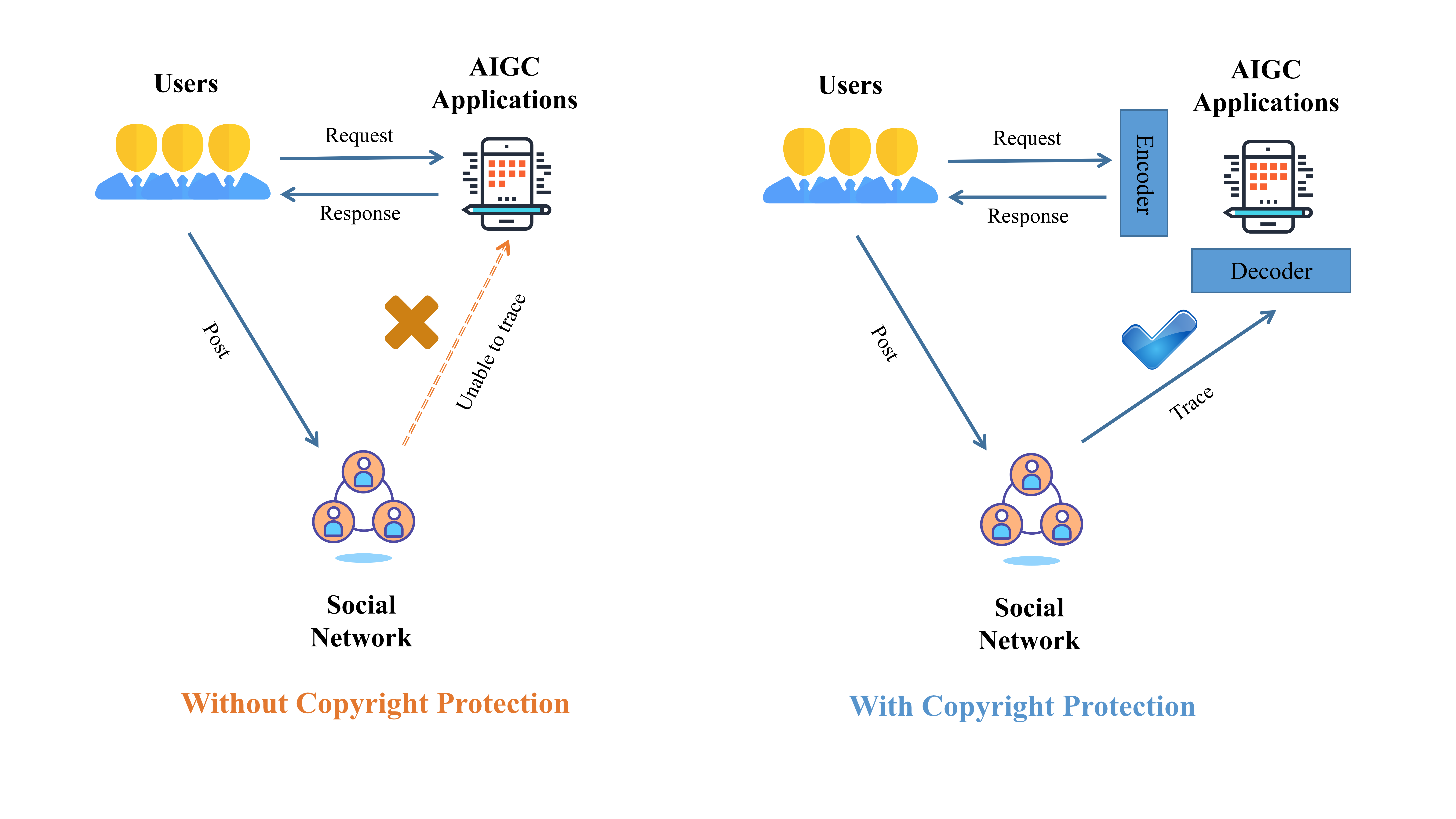}
	\caption{AIGC applications with (left) / without (right) copyright protection. The protection mechanism is the encoder-decoder framework with perturbation in \cite{wang2023plug}.}
	\label{fig: copyright_app}
\end{figure*}

\subsubsection{Tracing Attribution}

Generative network structures, such as GAN, are important tools in generative AI. The attribution of generative content also plays an important role in copyright protection. It allows IPR owners to differentiate between real images and fake images, and their source models by attributing fake images based on the architectural characteristics of the underlying model \cite{zhong2023copyright}. There are several methods for tracing the attribution of AIGC: spectral analysis, contrast attribution, and so on.

The method of spectral analysis proved to be effective in DeepFake detection and model attribution. Frank \emph{et al.} \cite{frank2020leveraging} proposed and validated detectors based on discrete cosine transformation (DCT). Their results reveal that in frequency space, GAN-generated images exhibit severe artifacts that can be easily identified. They demonstrate how frequency representation can be used to identify deep fake images in an automated way. This has important implications for content identification and copyright management for generative AI.

Contrast attribution is another method used to determine the modifications made by the GAN model to AIGC. This is done by analyzing the differences in local features between the original and generated contents. Yang \emph{et al.} \cite{yang2022deepfake} presented the first study on Deepfake Network Architecture Attribution to attribute fake images on architecture level. GAN architecture is likely to leave globally consistent fingerprints while traces left by model weights vary in different regions, and they  provide a simple yet effective solution named by DNA-Det for this problem.

\subsubsection{Ablating Concepts}The vast majority of copyright protection methods in the AIGC focus on the creator's perspective, preserving intellectual property by dealing with the original work. However, copyright is time-sensitive, and creators may wish to incorporate previously unprotected works into their intellectual property. What should be done if these works have already participated in the generation of AIGC at this time? Kumari \emph{et al.} \cite{kumari2023ablating} proposed an efficient method of ablating concepts in the pre-trained model, which prevents the generation of a target concept. This algorithm learns to match the image distribution for a target style, instance, or text prompt we wish to ablate to the distribution corresponding to an anchor concept. This prevents the model from generating target concepts given its text condition. Other similar work belongs to the domain of machine unlearning \cite{cao2015towards}, which explores scenarios in which users wish to delete specified data after model training. This is another possible direction for future AIGC copyright protection approaches.

\subsection{Defenses Against Adversarial Attacks}

The rapid development of AIGC cannot be achieved without the Deep Generative Model (DGM). Generative models synthesize data that we can observe in the world, and they generate content such as real-looking photos of human faces \cite{karras2017progressive}. This provides the technical basis for AIGC. As stated in Section III, these generative models are vulnerable to a variety of attacks. This puts the privacy and security of AIGC models at serious risk. We will discuss existing defense strategies that can secure the AIGC models in this subsection.

\subsubsection{Defenses Against Member Inference Attacks}"Membership" means whether a sample belongs to the training set of the machine learning model. And determining if the query sample follows the generated data distribution is called membership inference. In Membership Inference Attacks (MIA) \cite{shokri2017membership}, the attacker is given a data record and black-box query access to the target model. The attack succeeds if the attacker can correctly determine whether this data record was part of the model’s training dataset or not. The following solutions can be used to defend against membership inference attacks received by the AIGC model.

\begin{itemize}
    \item Weight Normalization. It is a reparameterization of the weight vectors in a neural network that decouples the length of those weight vectors from their direction. By reparameterizing the weights in this way, the conditioning of the optimization problem is improved, and the convergence of stochastic gradient descent is accelerated \cite{salimans2016weight}. This can also improve the generalization ability of generative models such as the AIGC model. The disadvantage is that it tends to lead to training instability \cite{hayes2017logan}.
    \item Dropout. Dropout \cite{srivastava2014dropout} is a technique for addressing overfitting. The key idea is to randomly drop units (along with their connections) from the neural network during training. This prevents units from co-adapting too much. Hayes \emph{et al.} \cite{hayes2017logan} applied dropout to DGM. However, they found that the dropout operation blurs the generated images and causes a slowdown in training. Therefore, more epochs may be required to obtain reliable generated samples in the case of dropout.
    \item Differentially-Private Stochastic Gradient Descent (DPSGD). Differential Privacy (DP) \cite{dwork2006differential} is a very effective privacy protection mechanism. DPSGD is an example of using DP for deep generative networks \cite{beaulieu2019privacy}. DPSGD uses a small amount of noise during the training process to slightly interfere with the optimization process. Because the discriminator has direct access to the training data, it is seen as key to achieving privacy preservation in GAN \cite{zhang2018differentially}. DPGAN \cite{xie2018differentially} and dp-GAN \cite{zhang2018differentially} have implemented the DPSGD method in the discriminator. One of them, DPGAN, limits the gradient by cropping the weights. And dp-GAN directly uses adaptive methods to crop the gradients. In GSWGAN \cite{chen2020gs},  only the gradient transferred from discriminator to generator follows the DPSGD method. DP-CGAN \cite{torkzadehmahani2019dp} is proposed for conditional GAN\cite{mirza2014conditional}. The method splits the loss of the discriminator between the real data and the generated data, and then crops the gradients of the two losses separately. These two are summed to obtain the overall gradient of the discriminator, and finally the noise is added to the overall gradient. Some studies have shown that DPSGD can defend against MIA in generative model \cite{chen2020gan}. However, such methods increase the computational complexity of the model, reduce the utility of the model, can affect the quality of the generated samples, and lead to longer training times.
    \item Change model architecture. As mentioned before, the MIA in DGM is based on the fact that the generated data distribution approximates the training data distribution. privGAN \cite{mukherjee2021privgan} is a new GAN architecture which tries to destroy the explicit approximation. It has multiple generator-discriminators. The training data are randomly partitioned into multiple partitions, and each partition is trained separately as a generator-discriminator pair. In this training mode, there are multiple approximate training data distributions and approximate generated data distributions, which interfere with the original approximation conditions and have some resistance to MIA. PATE-GAN \cite{jordon2019pate} is a framework which combines GANs with a Private Aggregation of Teacher Ensembles (PATE) to achieve  differential privacy guarantees. It trains a differential privacy discriminator so that the generator and the samples it generates have differential privacy guarantees.
    \item Expanding the training set. Extending the training set allows the model to cover more real samples and makes the training data distribution closer to the real training distribution. Data augmentation \cite{perez2017effectiveness} is usually used to extend the training data. A generalizable and balanced training set leads to a DGM that is more robust to MIA.
\end{itemize}

\subsubsection{Defenses Against Model Extractions Attacks}

\begin{table*}[h]
\centering
\renewcommand{\arraystretch}{1.3}
\caption{Overviews of Defenses Against Adversarial Attacks in AIGC models}
\resizebox{\columnwidth}{!}{
\begin{tabular}{|l|l|l|l|l|}
\hline
\textbf{Attack Type}                       & \textbf{Strategy}           & \textbf{Reference} & \textbf{Concept}                                                                                                                                                     & \textbf{Defect}                                                                                                                               \\ \hline
\multirow{5}{*}{Member inference attack}   & Weight normalization        &   \cite{salimans2016weight}                 & \begin{tabular}[c]{@{}l@{}}Reparameterize the weight \\ vectors in a neural network\end{tabular}                                                                     & Leads to training instability                                                                                                                 \\ \cline{2-5} 
                                           & Dropout                     &   \cite{hayes2017logan}                 & \begin{tabular}[c]{@{}l@{}}Randomly drop units from \\ the neural network during \\ training\end{tabular}                                                               & \begin{tabular}[c]{@{}l@{}}Blurs the generated images and \\ causes a slowdown in training\end{tabular}                                       \\ \cline{2-5} 
                                           & DPSGD                       & \cite{chen2020gan}                    & \begin{tabular}[c]{@{}l@{}}Uses differential privacy for \\ deep generative networks\end{tabular}                                                                    & \begin{tabular}[c]{@{}l@{}}Increase the computational \\ complexity  of the model and affects \\ the quality  of generated samples\end{tabular} \\ \cline{2-5} 
                                           & Model architecture(PrivGAN) & \cite{mukherjee2021privgan}                   & \begin{tabular}[c]{@{}l@{}}Destroy the approximation \\ between the training and \\ generated data\end{tabular}                                                         & \textbackslash{}                                                                                                                              \\ \cline{2-5} 
                                           & Model architecture(PATE-GAN) & \cite{jordon2019pate}                   & \begin{tabular}[c]{@{}l@{}}Employs PATE framework \\ to achieve differential \\ privacy\end{tabular}                                                         & \textbackslash{}                                                                                                                              \\ \cline{2-5}
                                           & Expanding the training set  &        \cite{perez2017effectiveness}            & \begin{tabular}[c]{@{}l@{}}Makes the training data \\ distribution closer to the \\ real training distribution, \\ such as data augmentation\end{tabular}               & Increases the training cost                                                                                                                   \\ \hline
\multirow{3}{*}{Model extractions attacks} & Digital watermarking        &         \cite{ong2021protecting}           & \begin{tabular}[c]{@{}l@{}}Verifies the ownership of \\ generative models by \\ embedding a digital watermark\end{tabular}                                              & \textbackslash{}                                                                                                                              \\ \cline{2-5} 
                                           & Output perturbation         & \cite{hu2021model}                   & \begin{tabular}[c]{@{}l@{}}Destroy the approximation \\ between the training and \\ generated data by adding \\ perturbation to generated samples\end{tabular}          & Degrades the image quality                                                                                                                    \\ \cline{2-5} 
                                           & Input perturbation          & \cite{hu2021model}   \cite{shen2020interpreting}                 & \begin{tabular}[c]{@{}l@{}}Destroy the approximation \\ between the  training and \\ generated data by adding \\ perturbation to training data\end{tabular}              & \textbackslash{}                                                                                                                              \\ \hline
\multirow{2}{*}{Evasion attacks}           & Smooth VAEs                 &    \cite{sun2020double} \cite{willetts2019improving}                & \begin{tabular}[c]{@{}l@{}}Smooth the mapping from \\ the input samples to the \\ latent code and mapping from \\ the latent code to the generated \\ samples\end{tabular} & \textbackslash{}                                                                                                                              \\ \cline{2-5} 
                                           & Model architecture(RoCGAN)  &        \cite{chrysos2020rocgan}            & \begin{tabular}[c]{@{}l@{}}Constrains the mapping from \\ the latent code to the generated \\ samples\end{tabular}                                                      & \textbackslash{}                                                                                                                              \\ \hline
\end{tabular}}
\end{table*}

The purpose of the model extraction attack is to replicate the target model by building a local model. While earlier such approaches mainly targeted deep discriminative models (DDM), more recently extraction attacks against deep generative models (DGM) are being explored. This also poses a threat to the security of AIGC models. Some existing approaches for defending against model extraction attacks for DGMs are discussed below.

\begin{itemize}
    \item Digital Watermarking. Digital watermarking does not prevent model theft but provides proof at the intellectual property level. Thus using digital watermarking to embed identifying information into network parameters can be a defense against model extraction attacks. Ong \emph{et al.} \cite{ong2021protecting} first proposed to protect the intellectual property of GAN with digital watermarking technique. The method trains the model to generate samples with a specific identity when a specific tag is an input. They also use sign loss \cite{fan2019rethinking} to embed the identification information into a normalization layer in the generator for ownership verification.
    \item Output Perturbation. Due to the similarity between the distribution of the training data and the distribution of the generated data, the private data of the AIGC model and the training set are vulnerable to theft. For this reason, the generated samples can be perturbed to destroy the similarity between them. Hu \emph{et al.} \cite{hu2021model} evaluated the effects of four perturbation methods. Adding Gaussian distributed additive noise, adding adversarial noise to ensure the perturbed image would be misclassified, Gaussian filtering, and JPEG compression. The experimental results indicate that the most stable defense method is to add Gaussian noise, but this degrades the image quality.
    \item Input Perturbation. Another way to destroy the similarity between the training and generated data is to perturb the input data and thus the generated samples. Hu \emph{et al.} \cite{hu2021model} described two methods, linear interpolation and semantic interpolation. Linear interpolation extracts several interpolated latent points between two samples and uses them as model inputs. This can cause disturbances in the distribution of the generated data. Shen \emph{et al.} \cite{shen2020interpreting} proposed a novel framework, called InterFaceGAN, for semantic face editing by interpreting the latent semantics learned by GANs. Hu adopted the semantic interpolation algorithm proposed by Shen and used each approach to defend against their proposed model extraction attack. The results showed semantic interpolation to have a more stable and more effective performance \cite{sun2021adversarial}.

\end{itemize}

\subsubsection{Defenses Against Evasion Attacks}Evasion attacks lead to unsatisfactory outputs by processing the model inputs. For the deep generative model (DGM) used for AIGC, the model inputs include latent codes and input samples. Several methods that can be used to defend the AIGC model against evasion attacks are discussed below.

\begin{itemize}
    \item Smooth VAEs. As an important technical support for generative networks, VAEs are vulnerable to adversarial samples. Even a small change in the input data can lead to significant changes in the latent distribution derived from the input. Therefore, an effective way to defend against evasion attacks is to mitigate such changes, i.e., smoothness. Sun \emph{et al.} \cite{sun2020double} used double backpropagation to achieve smoothing of the VAEs. This approach limits the gradient of the reconstructed image to the original image, making the autoencoder insensitive to any small perturbations inserted as part of the attack. Willetts \emph{et al.} \cite{willetts2019improving} proposed disentangled representation, which provides another way of thinking to achieve smoothing. This method regularizes the network by penalizing the total correlation (TC) term. The total correlation term quantifies the degree of dependence between different latent dimensions in the aggregated posterior, thus allowing the aggregated posterior to factorize across dimensions. VAE with TC penalty is not only more robust to adversarial attacks, but also provides better reconstruction performance.
    \item Change model architecture. Conditional GANs (cGAN) \cite{mirza2014conditional} can generate samples conditioned on labels by providing additional labels. cGAN is not explicitly constrained to model outputs, therefore, it is very vulnerable to adversarial inputs, i.e., evasion attacks. Chrysos \emph{et al.} \cite{chrysos2020rocgan} proposed robust conditional GAN (RoCGAN). RoCGAN uses an additional unsupervised mapping process called the AE pathway, which is distinguished from the traditional supervised mapping process called the reg pathway. Both of them work like an encoder. RoCGAN shares decoder weights in both paths to force the latent representations of them to be semantically similar, which limits the output of the reg path.
\end{itemize}

\section{Open Issues and Future Directions}
The issue of privacy and security in AIGC has attracted much attention. Although there have been many solutions from different perspectives, how to defend the data security of AIGC models is still an area that is far from being fully explored. We outlook the future direction of AIGC privacy and security from several perspectives, such as new scenarios, possible technologies and so on.

\subsection{High-risk Scenarios}
The AIGC model has already played an important role in a variety of areas of human society, and the quality of the text, images, audio and video it generates has reached a certain level. However, currently, AIGC serves more in the field of higher fault tolerance. On the contrary, the use of AIGC in some high-risk scenarios is still very immature, such as healthcare \cite{reddy2020governance}, finance \cite{qi2018fintech}, autonomous driving \cite{grigorescu2020survey} and scientific research \cite{gil2014amplify}. The use of AIGC in these fields is often related to the safety of human life, so there are high requirements for the quality and safety of AIGC and very low fault tolerance. Research still has a long way to go to improve the security of AIGC to a level where we can use it with confidence.

\subsection{Timeliness}
In order to ensure that AIGC accurately reflects the current state of society, its training corpus must be updated regularly. This also prevents the information lag of AIGC and ensures that the model is kept up to date. Lazaridou \emph{et al.} \cite{lazaridou2021pitfalls} pointed out that the transformer model cannot accurately predict data that is not part of the training data period. This is because the test data and training data belong to different periods, and increasing the size of the model does not solve this problem. Therefore, it is necessary to collect new training data in the AIGC model and update the model periodically in order to make its output time-sensitive.  

\subsection{Sustainability}
In view of enabling AIGC to serve users in real-time, the study of mobile AIGC networks is crucial. The life cycle of generative models contains processes such as pre-training, fine-tuning and inference, all of which consume significant computational and communication resources in the network \cite{mao2017survey}. For this reason, efficient algorithms and frameworks should be developed to utilize intelligent resource management and scheduling techniques to enable AIGC networks to operate smoothly and continuously despite the dynamic service configuration and the operating mode of edge nodes. Failure to ensure the sustainability of the AIGC network may lead to the collapse of the network, further affecting the quality of generated content and threatening the security of users.

\subsection{Fairness}To ensure the safety of AIGC, we have to consider the differential impact of AIGC on different populations, i.e., fairness. A poorly fair AIGC model may further exacerbate global inequality \cite{weidinger2021ethical}. How to equitably distribute the benefits of AIGC models is an open and worthy question to explore. 

\subsection{AIGC Meets Emerging Fields}The rapid development of artificial intelligence has given rise to many emerging fields and concepts, which will inevitably intermingle with AIGC and trigger more thoughts on AIGC privacy and security. Representative fields include Web 3.0 and metaverse, etc. Web 3.0 has attracted considerable attention for its unique decentralized features and has become an important tool for empowering the digital economy. In the digital economy scenario, the centralized nature of the Internet and other features usually bring security issues such as infringement and privacy leakage. Chen \emph{et al.} \cite{chen2022digital} studied the latest advances in Web 3.0 for machine learning, finance and data management. However, the rise of AIGC has greatly enriched the Web 3.0-driven digital economy, including the content and manner of transactions. Therefore, it is necessary to reconsider the privacy and security issues in Web 3.0 scenarios that include AIGC.

\section{Conclusion}
As large model-driven AIGC services such as ChatGPT enter people's daily life, the privacy and security issues they bring to users gradually come to the fore and become one of the urgent issues to be solved in the AIGC era. We first provide a brief introduction to the definition, classification, and general techniques of AIGC, while emphasizing the urgency of protecting the privacy and security of AIGC, pointing out that techniques such as privacy computing can be combined with the AIGC model. We then describe the current challenges that AIGC faces from several perspectives, including the privacy of circulating data, the security of generated content, the copyright, and the threat of malicious users. We focused our research on solutions for privacy and security in AIGC, using technologies such as blockchain, federated learning, digital watermarking, differential privacy, and so on. They provide rich solutions to address the privacy and security issues of generative models including AIGC tools. However, the current solutions to privacy and security issues in AIGC are not mature enough, and the rapid development of large models has brought many new challenges to the AI field. We discuss some open issues and possible future directions in AIGC privacy and security at the end of the paper. We hope that this review provides an overview of privacy and security issues in AIGC and offers new ideas on how academia and industry can better utilize AIGC.

\bibliographystyle{ACM-Reference-Format}
\bibliography{sample-base}

\appendix

\end{document}